\documentclass[12pt, notitlepage]{article}
\usepackage[T1]{fontenc}
\usepackage[utf8]{inputenc}

\usepackage{authblk}

\usepackage{amsmath,amssymb,amsfonts}
\usepackage{graphicx}
\usepackage{bm}
\usepackage{color}
\usepackage{epstopdf,epsfig}
\usepackage[authoryear]{natbib}
\setcitestyle{round,aysep={},yysep={;}}
\usepackage{newtxtext}

\usepackage{newtxmath}

\usepackage{hyperref}
\hypersetup{
    colorlinks = true,
    urlcolor   = blue,
    citecolor  = blue,
}

\usepackage[letterpaper, left=1in, right=1in, bottom=1in, top=0.75in]{geometry}

\usepackage{orcidlink}

\numberwithin{equation}{section}

\usepackage{lipsum}

\newcommand{\bnabla}{\boldsymbol{\nabla}}
\newcommand{\bcdot}{\boldsymbol{\cdot}}

\newcommand{\be}{\boldsymbol{e}}
\newcommand{\bq}{\boldsymbol{q}}
\newcommand{\bn}{\boldsymbol{n}}
\newcommand{\bj}{\boldsymbol{j}}

\newcommand{\bx}{\boldsymbol{x}}
\newcommand{\bu}{\boldsymbol{u}}
\newcommand{\bU}{\boldsymbol{U}}
\newcommand{\bH}{\boldsymbol{H}}
\newcommand{\bomega}{\bm{\omega}}
\newcommand{\bOmega}{\bm{\varOmega}}
\newcommand{\bF}{\boldsymbol{F}}
\newcommand{\bL}{\boldsymbol{L}}
\newcommand{\boldm}{\boldsymbol{m}}

\newcommand{\bI}{\boldsymbol{I}}
\newcommand{\bK}{\boldsymbol{K}}
\newcommand{\bR}{\boldsymbol{R}}
\newcommand{\bsigma}{\boldsymbol{\sigma}}
\newcommand{\btau}{\boldsymbol{\tau}}
\newcommand{\bQ}{\boldsymbol{Q} }

\newcommand{\nb}{\overline{n}}

\usepackage{xcolor}

\newcommand{\ksts}{k_sT_s}
\newcommand{\kt}{k_BT}
\newcommand{\Pes}{Pe_s}

\begin{document}

\title{Activity-induced propulsion of a vesicle}
\date{\today}

\author[1]{Zhiwei Peng\orcidlink{0000-0002-9486-2837} }
\author[2]{Tingtao Zhou\orcidlink{0000-0002-1766-719X}}
\author[1,2]{John F. Brady\orcidlink{0000-0001-5817-9128}\thanks{Electronic mail: jfbrady@caltech.edu}}
\affil[1]{Division of Chemistry and Chemical Engineering, California Institute of Technology, Pasadena, California 91125, USA}
\affil[2]{Division of Engineering and Applied Science, California Institute of Technology, Pasadena, California 91125, USA}

\maketitle

\begin{abstract}
    Modern biomedical applications such as targeted drug delivery require a delivery system capable of enhanced transport beyond that of passive Brownian diffusion. In this work an osmotic mechanism for the propulsion of a vesicle immersed in a viscous fluid is proposed.  By maintaining a steady-state solute gradient inside the vesicle, a seepage flow of the solvent (e.g., water) across the semipermeable membrane is generated which in turn propels the vesicle. We develop a theoretical model for this vesicle-solute system in which the seepage flow is described by a Darcy flow. Using the reciprocal theorem for Stokes flow it is shown that the seepage velocity at the exterior surface of the vesicle generates a thrust force which is balanced by the hydrodynamic drag such that there is no net force on the vesicle. We characterize the motility of the vesicle in relation to the concentration distribution of the solute confined inside the vesicle.  Any osmotic solute is able to propel the vesicle so long as a concentration gradient is present. In the present work, we propose active Brownian particles (ABPs) as a solute. To maintain a symmetry-breaking concentration gradient, we consider ABPs with spatially varying swim speed and ABPs with constant properties but under the influence of an orienting field. In particular, it is shown that at high activity the vesicle velocity is $\bU\sim [K_\perp /(\eta_e\ell_m) ]\int \Pi_0^\mathrm{swim} \bn d\Omega $, where $\Pi_0^\mathrm{swim}$ is the swim pressure just outside the thin accumulation boundary layer on the vesicle interior surface, $\bn$ is the unit normal vector of the vesicle boundary, $K_\perp$ is the membrane permeability, $\eta_e$ is the viscosity of the solvent, and $\ell_m$ is the membrane thickness.
\end{abstract}

\section{Introduction}
\label{sec:intro}
Targeted drug delivery is an important goal of modern nanomedicine. Recent advances in the design, manufacture and control of nanocarriers have enabled the delivery of such cargoes into single cells for the purpose of imaging, diagnostics and therapeutics \citep{West2003, Gao2005,Rao2007,Torchelin2012}. Commonly used pharmaceutical nanocarriers include liposomes, micelles, nanoemulsions, polymeric nanoparticles and many others \citep{Torchelin2012}. In particular, liposomes have become an important class of carriers for the encapsulation and transport of medical cargoes because of several advantages including their biocompatibility with human cells, the improved solubility of drugs and versatility for chemical targeting \citep{Torchilin2015}, among others.

A liposome is a vesicle that has an aqueous solution core encircled by a hydrophobic membrane (lipid bilayer); hydrophilic solutes dissolved in the core cannot readily pass through the membrane while lipophilic chemicals tend to associate with the bilayer. As a result, a liposome can be loaded with hydrophilic, lipophilic and/or amphiphilic cargoes in the context of drug delivery. Recently, the Moderna vaccine developed to prevent coronavirus disease 2019
(COVID-19) has utilized a lipid based nanovesicle to encapsulate the mRNA vaccine that encodes the SARS-CoV-2 spike glycoprotein \citep{Jackson2020}.

The liposome-encapsulated medical cargo is transported passively, either via diffusion or advection due to local fluid flow, which limits its ability to overcome biological barriers. To mitigate such limitations of passive drug delivery, active drug delivery platforms using motile microrobots (or microswimmers), either synthetic or biohybrid, have been proposed \citep{Medina2018, Erkoc2019,Singh2019,Bunea2020}. By attaching nanoparticle cargoes to the surface of a motile microswimmer, the delivery system can actively navigate, access regions that are unreachable to passive drug delivery, and be directed to the desired site using chemotaxis or an external magnetic field \citep{felfoul2016magneto,Park2017}. Due to self-propulsion of the microswimmer, the effective dispersion of the attached cargo is greatly enhanced, sometimes by a few orders of magnitude, compared to the long-time self diffusivity of the passively-transported cargo \citep{Singh2017}.

Instead of attaching a cargo to the surface of a microswimmer, one can also encapsulate both the cargo and the microswimmer inside the vesicle. Encapsulated microswimmers have been studied by previous works. For example, biological microswimmers and self-propelled Janus particles haven been successfully encapsulated inside engineered giant unilamellar  vesicles (GUVs) \citep{Trantidou2018,Takatori2020,vutukuri2020active}.
The encapsulated microswimmer provides the vesicle with enhanced super-diffusive motion mediated through hydrodynamic interactions between the microswimmer and the vesicle provided that the fluid is allowed to pass through the membrane of the vesicle~\citep{marshall_brady_2021}.

In the present work we consider a system that combines the benefits of the vesicle for cargo encapsulation and the self-propulsion of microswimmers for enhanced transport.
We propose an alternate model system in which the vesicle is propelled by an osmotic flow that is induced by an actively-maintained concentration gradient of a solute inside the vesicle. This kind of osmotic propulsion has been proposed as an alternate mechanism for tumor cells to migrate under strong confinement, in which case other modes of motility such as contractility is inhibited. \citet{STROKA2014611} showed that through physical and biochemical processes, the tumor cell establishes a spatial gradient of solute (ions), which creates a net inflow of water at the cell leading edge and a net outflow at the cell trailing edge. As a result, this water permeation process enables the cell to migrate through narrow channels. We are specifically interested in studying the motility of the vesicle as a result of a prescribed concentration gradient of a solute that is confined inside the vesicle. Because the solute particles are not allowed to pass through the membrane, an osmotic flow of water is generated, which in turn propels the vesicle immersed in water.

The main question we wish to address in this work is: What is the motility of the vesicle system in relation to the concentration gradient of the solute? More interestingly, does the vesicle move in the same or opposite direction of the concentration gradient?

We show by explicit calculation that for a weakly permeable membrane the translational velocity of a rigid spherical vesicle becomes
\begin{equation}
    \label{eq:intro-U-Da0}
        \bU = \frac{1}{4\pi} \frac{K_\perp}{\eta_e \ell_m} \int_{S^2} \Pi^\mathrm{osmo}_0 \bn d \Omega,
\end{equation}
where $\Pi_0^\mathrm{osmo} = n^w k_BT$ is the osmotic pressure of the solute at the interior wall, $n^w$ is the local number density of the solute in the absence of internal fluid flow, $k_BT$ is the thermal energy, $K_\perp$ is the membrane permeability, $\eta_e$ is the viscosity of the solvent (water) and $\ell_m$ is the thickness of the membrane. In equation \eqref{eq:intro-U-Da0}, $\bn$ is the unit outward normal vector (see figure \ref{fig:problem-schematic}) and the integration is over the solid angle in three dimensions (3D). In this limit, the translational velocity of the vesicle is linearly proportional to the driving force---the osmotic pressure. As expected, a number density at the interior wall that breaks front-back symmetry is required in order to have a nonzero translational velocity of the vesicle.

Equation \eqref{eq:intro-U-Da0} applies generally for any osmotic solute in the weak permeability limit so that the interior fluid flow only slightly perturbs the solute distribution. For example, a linear solute gradient, $n_0 = n_0(\bm{0}) + \bx\bcdot \bnabla n_0$, results in
\begin{equation}
    \label{eq:intro-linear-grad}
    \bU = \frac{1}{3} \frac{K_\perp}{\eta_e \ell_m} (R-\ell_m) k_BT \bnabla n_0,
\end{equation}
where $\bnabla n_0$ is a constant vector and $R$ is the exterior radius of the vesicle. Therefore, for the simple prescribed linear-density gradient, the vesicle translates in the same direction as the gradient in number density.

The above discussion reveals that the vesicle is able to exhibit net motion when an interior solute concentration gradient is given. A separate, but important, question is: How can such a solute gradient be maintained? For a biological cell, this is achieved by its internal physical and biochemical processes \citep{STROKA2014611}. For a synthetic vesicle system for the purpose of enhanced transport, alternate methods need to be implemented in order to generate such a concentration gradient.

In this work, leveraging recent advances in the understanding of the dynamics of active matter, we propose to use active Brownian particles (ABPs) as the solute. In addition to normal thermal Brownian motion with translational diffusivity $D_T$, ABPs self-propel with an intrinsic  `swim' speed $U_s$ in a direction $\bq$. The orientation of the swimming direction $\bq$ changes on a reorientation timescale $\tau_R$ that results from either continuous random Brownian rotations or the often-observed discrete tumbling events of bacteria. One important intrinsic length scale due to activity is the run or persistence length $\ell = U_s\tau_R$. Previous works have shown that a spatial variation in the swim speed leads to a spatial variation in the concentration (or number density) of active particles \citep{Schnitzer1993, Tailleur2008,Row2020}. By tuning the swim speed distribution of ABPs confined inside the vesicle, a spherically asymmetric density distribution can emerge and lead to net motion of the vesicle.

For active particles with slow spatial variation in swim speed in 1D, \citet{Schnitzer1993} and later \citet{Tailleur2008} showed that the local number density $n$ is inversely proportional to the local swim speed $U_s$, i.e., $nU_s = const$. This simple prediction has been validated experimentally using bacteria that swim with an intensity-dependent speed when illuminated by a spatial light pattern \citep{arlt2019dynamics}. \citet{Row2020} generalized this result and showed that the spatial variation in activity (e.g., swim speed) can be utilized as a pump mechanism in which fluid flows from regions of high concentration of particles to low. Employing this spatial variation, we show that encapsulated ABPs with spatially varying activity can be used to propel the vesicle.

In equations \eqref{eq:intro-U-Da0} and \eqref{eq:intro-linear-grad}, the vesicle velocity appears to be linearly proportional to $\kt$. However, this does not imply that the driving force is necessarily thermal in origin (in thermodynamic equilibrium no density gradient is present). In the case of ABPs as solute, the active (non-equilibrium) dynamics provides such a density gradient. Analogous to the Stokes-Einstein-Sutherland relation $\kt = \zeta D_T$, where $\zeta$ is the Stokes drag coefficient, an active energy scale $\ksts = \zeta \tilde{D}^\mathrm{swim}$ can be defined for active matter systems \citep{TakatorietalPRL}, where $ \tilde{D}^\mathrm{swim} = \tilde{U}_s^2 \tau_R/6$ is the swim diffusivity. We note that for ABPs with spatially varying swim speed a characteristic swim speed $\tilde{U}_s$  is used in the definition of the swim diffusivity; the local active energy $\ksts(\bx)$ can also be defined by using the local swim speed $U_s(\bx)$ and/or local reorientation time $\tau_R(\bx)$. An important parameter that quantifies the activity of ABPs is the ratio $\ksts/\kt = \tilde{D}^\mathrm{swim}/D_T$. For many active matter systems this ratio is very large, often exceeding $10^3$ \citep{takatori2016acoustic}. In this high activity limit, the ABPs exhibit a thin accumulation boundary layer at the interior surface of the vesicle. As we shall show in section \ref{sec:high-activity-Da}, the local density at the interior wall of the vesicle can be related to the density just outside the boundary layer via the equation $n^w\kt=n^0 \ksts(\bx)f=\Pi_0^\mathrm{swim}(\bx)f$, where $\Pi_0^\mathrm{swim}$ is the swim pressure just outside the boundary layer and $f$ is a factor that depends on the ratio of the run length to the size of the vesicle. [ This factor is unity for the case of ABPs on one side of an infinite planar wall \citep{yan_brady_2015}.] For highly active ($\ksts \gg \kt$) ABPs, equation \eqref{eq:intro-U-Da0} becomes
\begin{equation}
    \label{eq:speed-high-activity}
\bU = \frac{1}{4\pi} \frac{K_\perp}{\eta_e\ell_m}  \int_{S^2} n^0 \ksts(\bx) f\bn d \Omega = \frac{1}{4\pi} \frac{K_\perp}{\eta_e\ell_m}  \int_{S^2} \Pi_0^\mathrm{swim} \bn f d \Omega,
\end{equation}
showing that the velocity of the vesicle is proportional to the swim pressure. More precisely, it is the variation of the swim pressure [due to the variation in swim speed or run length $\ell(\bx$)] that gives rise to net motion.

Instead of using ABPs with spatially varying swim speed or run length, one can also consider using an external field that orients constant-property ABPs towards a certain direction. External fields such as chemical gradients or magnetic fields can affect the swimming behavior of microorganisms to facilitate their  movement towards a favorable region. In the laboratory, an externally applied magnetic field has been used to guide nanocarriers for the purpose of targeted drug delivery \citep{felfoul2016magneto, Torchilin2015}. In the presence of an external orienting field, even for ABPs with constant properties, the front-back symmetry is broken, and net motion of the vesicle is generated. The balance of the strength of the orienting field and the random reorientation due to rotary diffusion is characterized by the Langevin parameter, $\chi_R=\Omega_c \tau_R$, where $\Omega_c$ is the strength of the angular velocity induced by the field \citep{TakatoriSM}. Noting that the force exerted by the active particles on the wall $\bF^w = \kt \int n^w \bn dS $ \citep{yan_brady_2015}, we rewrite equation \eqref{eq:intro-U-Da0} as $\bU = K_\perp \bF^W /(4\pi R^2\eta_e \ell_m)$. In other words, we need to know the net force the active particles exert on the wall to determine the net vesicle motion. The force on the wall scales as $N^w \zeta U_s$, where $N^w$ is the total number of particles at the wall and each particle pushes against the wall with at most its swim force $\zeta U_s$. The balance of this force due to the ABPs with the drag force of the porous vesicle moving through an external viscous fluid gives the net motion. Of particular interest is the strong-field limit, where the number of particles on the wall is on the same order as the total number of particles, $N^w/N =O(1)$, and the net speed of the vesicle is the largest, $U \sim K_\perp N \zeta U_s /(R^2 \eta_e\ell_m)$.

This last example where we argued that the vesicle motion can be deduced from the net swim force of the ABPs balancing the drag of the vesicle also applies to the so-called `dry' active matter \citep{Marchetti_review}. Dry active matter describes bacteria (or other organisms) that crawl (or even walk) on a surface of a medium of resistivity $\zeta$. Active particles confined to a `container' that is able to slide along the surface in response to a lateral force will be able to push the container via their `swim' force if there is an asymmetric distribution of ABPs. The net swim force would scale as $N^w\zeta U_s$, and the container would translate with the speed $U_c\sim N^w\zeta U_s/\zeta_c$, where $\zeta_c$ is the resistivity for sliding the container along the surface.
For dry active matter there is no fluid and thus one does not have the notion of a semipermeable membrane nor a seepage velocity driven by an osmotic pressure difference. Nevertheless, the mechanics are the same: like the seepage velocity, the substrate surface must move across the container boundary as it slides along the surface, and the ABPs achieve their propulsive `crawling' force by pushing off the substrate just like swimmers push off the fluid. Thus, at least at high activity, the results derived here apply equally well to dry active matter with an appropriate change in notation.

In the case of a spherical vesicle, its net motion is induced by an asymmetric number density distribution on the vesicle interior surface. An alternate route for the generation of net motion is to use a vesicle with an asymmetric shape. Because the accumulation of ABPs at the interior surface depends on the local curvature of the boundary, a vesicle that has a front-back asymmetry in its shape is able to exhibit net motion. Indeed, the exterior version of the problem where a passive object is immersed in a bath of active particles has been studied. It has been shown in experiments and simulations that for an object with shape asymmetry, net motion can be achieved \citep{Sokolov969, Kaiser2014,Yan2018}.

To obtain the results for the vesicle motility, in section \ref{sec:formulation} we describe the model and derive a theoretical formulation that governs the dynamics of the vesicle, the interior solute  suspension  and the exterior fluid flow. A Darcy-like constitutive law that models the response of the fluid seepage velocity in relation to the fluid stress differences across the membrane is used. This formulation is at the continuum level, where the vesicle is large compared to the size of the ABPs so that the interior (fluid and ABPs) is treated as a suspension; the suspension stress includes the fluid stress and the osmotic pressure of the ABPs. The exterior flow field satisfies the boundary condition that the fluid velocity at the exterior surface of the vesicle consists of the rigid body motion and a seepage velocity. Because the vesicle is force- and torque-free, we can relate the rigid body motion to the seepage velocity distribution at the exterior surface using the reciprocal theorem. This approach is similar to treatments of the swimming of microorganisms using the squirmer model \citep{Stone1996} where the boundary velocity at the surface of the swimmer is decomposed into rigid-body motion and the slip velocity distribution.

In situations relevant for the vesicle model as we consider here, the interior fluid flow is often weak compared to the active self-propulsion. In section \ref{sec:weak-interior-flow}, by neglecting the interior fluid flow we show that the total (fluid and osmotic) pressure inside the vesicle is constant and the leading-order translational velocity of the vesicle is driven by the difference in the fluid pressure across the membrane. As a result, one only needs to compute the distribution of ABPs in the absence of flow and the resulting number density distribution at the interior wall is used to obtain the translational velocity. The effect of an external orienting field on the dynamics of confined ABPs and the motion of the vesicle is considered in section \ref{sec:external-field}. The behavior of ABPs with slow spatial variation in their swim speed where fluid motion is explicitly considered is discussed in section \ref{sec:slow-activity-variation}. Finally, we conclude in section \ref{sec:conclusion} with a discussion of the limitations and extensions of this vesicle-ABPs propulsion system.

\section{Problem formulation}
\label{sec:formulation}
\begin{figure}
  \centering
  \includegraphics[scale=1.0]{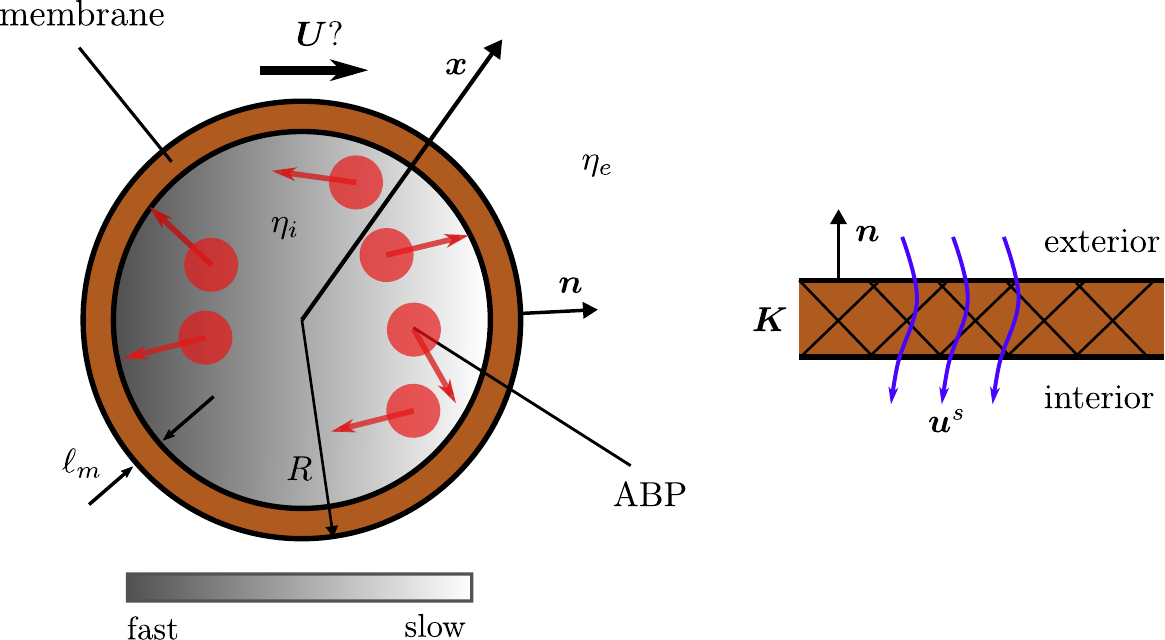}
  \caption{\label{fig:problem-schematic}Left: A rigid spherical vesicle with a semipermeable membrane immersed in an otherwise quiescent viscous fluid. Active Brownian particles are confined inside the vesicle. Right: Schematic of the semipermeable membrane with a permeability tensor $\bK$ and thickness $\ell_m$. The seepage velocity in the membrane is $\bu^s$, which in general depends on the local position vector.}
\end{figure}

Consider a rigid vesicle or cell consisting of a thin membrane and a solution core immersed in an otherwise quiescent viscous fluid (see figure~\ref{fig:problem-schematic}). The interior of the vesicle is a suspension of potentially active elements, which we model as active Brownian particles. The boundary or membrane of the vesicle is permeable to the solvent (i.e. water) but not to the solute (ABPs). In other words, the membrane is an osmotic membrane and serves as a confining boundary for the ABPs. Relative to the vesicle, the fluid domain is partitioned into interior, exterior and the thin porous (in the membrane) regions. The solvent in all regions is identical.

The ABPs encapsulated inside the vesicle swim with a prescribed spatially varying swim speed, which is the driving mechanism for a spatially varying number density.

At small scales relevant to the vesicle-ABP system proposed here, the inertia of the fluid, the ABPs and the vesicle are negligible. In particular, for motile bacteria such as \textit{E. coli}, which has a characteristic size of $\sim$1\textmu m and a swim speed of $\sim$30\textmu m/s, the Reynolds number in water is $3 \times 10^{-5}$. The resulting speed of the vesicle and the Reynolds number based on the size of the vesicle and its speed are also small. In this low Reynolds number limit, the dynamics of the fluid is governed by the Stokes equations and there is no external force/torque on the vesicle.

\subsection{The exterior flow}
The exterior domain consists of solvent alone and its dynamics is governed by
\begin{equation}
     \bnabla\bcdot\bsigma_f^e = \eta_e \nabla^2 \bu^e - \bnabla p_f^e =\boldsymbol{0}, \quad \bnabla\bcdot\bu^e=0.
\end{equation}
Here, $\bsigma_f^e$ is the stress tensor, $\eta_e$ is the dynamic viscosity of the solvent, $p_f^e$ is the pressure field and $\bu^e$ is the velocity field. Far from the vesicle, the fluid is undisturbed and there is no background flow:
\begin{equation}
    p_f^e \to 0 \quad\mbox{and}\quad \bu^e \to \bm{0}\quad\mbox{as}\quad r \to \infty.
\end{equation}
At the exterior surface of the vesicle, we have
\begin{equation}\label{eq:exterior-bc}
    \bu^e\left(\bx \in S_e\right) = \bU +\bOmega \times \bx + \bu^s(\bx),
\end{equation}
where $S_e$ denotes the exterior surface of the vesicle, $\bU$ ($\bOmega$) is the rigid-body linear (angular) velocity of the vesicle and $\bu^s$ is the local seepage velocity at the exterior surface. The definition of $\bu^s$ is deferred to Section \ref{sec:membrane}. We note that equation \eqref{eq:exterior-bc} is similar to the squirmer model where the closely packed cilia tips of a microorganism are modeled as a distribution of radial and tangential velocities on the cell body, often taken to be of spherical shape \citep{lighthill1952squirming,blake_1971}.

\subsection{The interior suspension}
The particles and solvent in the interior of the vesicle are treated as a continuum and governed by
\begin{equation}
    \bnabla\bcdot\bsigma^i = \eta_i \nabla^2 \bu^i - \bnabla P =\boldsymbol{0} \quad\mbox{and}\quad \bnabla\bcdot\bu^i = 0,
\end{equation}
where $\bsigma^i$ is the stress tensor, $\eta_i$ is the dynamic viscosity of the suspension and $\bu^i$ is the velocity field. Here, the total pressure is given by
\begin{equation}
 P = p_f^i + n \kt,
\end{equation}
where $p_f^i$ is the fluid pressure, $n$ the number density of the ABPs and $\kt$ is the thermal energy. In our model, the only  contribution to the suspension stress from the ABPs is the osmotic pressure $n\kt$.

Here, the swim pressure introduced by \citet{TakatorietalPRL} does not directly enter the analysis. Regardless of activity, the particle contribution to the stress is $\bsigma_p = - n k_BT \bI$. In the high activity limit, however, as shown in equation \eqref{eq:speed-high-activity},  the vesicle motion ultimately results from the swim pressure variation. Furthermore, we note that additional stress contributions such as the active hydrodynamic stresslet of ABPs~\citep{Saintillan2015} can be readily incorporated into our model. Since the osmotic pressure is  present regardless of activity, in this paper we focus on the osmotic pressure and neglect additional stress contributions.

At the interior wall of the vesicle, we have
\begin{equation}
    \bu^i\left(\bx\in S_i\right) = \bU +\bOmega \times \bx + \bu^s(\bx),
\end{equation}
where $S_i$ is the interior surface of the vesicle.

\subsection{Dynamics of ABPs}
The distribution of ABPs confined inside the vesicle is described by the probability density $\varPsi(\bx, \bq, t)$ as a function of space $\bx$, orientation $\bq$ ($|\bq|=1$) and time $t$. The conservation of ABPs is governed by the Smoluchowski equation. At steady state, this is given by
\begin{equation}\label{eq:smol}
    \bnabla\bcdot\bj_T + \bnabla_R\bcdot\bj_R =0,
\end{equation}
where the translational and rotational fluxes are given by, respectively,
\begin{eqnarray}
    \bj_T&=& \bu^i \varPsi + U_s(\bx) \bq \varPsi -D_T\bnabla \varPsi, \\
    \bj_R&=& \frac{1}{2}\bomega^i \varPsi - D_R\bnabla_R \varPsi.
\end{eqnarray}
Here, $D_T$ is the thermal diffusivity of ABPs, $\bomega^i = \bnabla\times\bu^i$ is the vorticity vector, $D_R$ is the rotary diffusivity, $\bnabla_R = \bq\times\bnabla_q$ is the rotary gradient operator and $U_s(\bx)$ is the intrinsic swim speed of ABPs. The prescribed spatial variation of $U_s$ is the key ingredient of our model, and is responsible for the generation of a concentration gradient of ABPs inside the vesicle.

The conservation of ABPs requires that
\begin{equation}
    \int_{V_i} n  d\bx  = N,
\end{equation}
where $n = \int \Psi d\bq$ is the number density, $N$ is the total number of ABPs and $V_i$ is the volume of the interior of the vesicle. At the interior surface of the vesicle, the flux relative to the rigid-body motion must vanish. This no-flux condition can be written as
\begin{equation}\label{eq:noflux}
    \bn \bcdot\bj_T = \bn\bcdot\left(\bU+ \bOmega\times \bx \right)\varPsi,\quad \bx \in S_i,
\end{equation}
where $\bn$ is the unit normal vector as shown in figure \ref{fig:problem-schematic}. We note that as a model of active elements inside a cell, the rotary diffusivity $D_R$ is biological rather than thermal in origin. As a result, $D_R$ is independent of $D_T$ (which is assumed to be thermal in origin). The rotary diffusivity defines a reorientation timescale $\tau_R = 1/D_R$ that characterizes the relaxation of the swimming direction. The ABPs take a step of magnitude $\ell = U_s\tau_R$, which is often called the run (or persistence) length $\ell$,  before its swimming direction changes significantly. Note that one might have a reorientation time $\tau_R(\bx)$ that is a function of position in addition to a spatially varying swim speed, as we show below that the important quantity is the run length $\ell(\bx)$.

In contrast to passive Brownian particles, the self-propulsion of active particles introduces a coupling between their rotational and translational dynamics via the swimming motion. That is, even for an isolated active Brownian sphere (which is geometrically isotropic), one must track both its orientation and position. One manifestation of such a coupling is the enhanced long-time self-diffusivity beyond the thermal diffusivity $D_T$, which for an ABP with constant properties in free space is $D^\text{eff} = D_T + D^\text{swim}$, where $D^\text{swim} = U_s^2\tau_R/6$ (in 3D) is the swim diffusivity. In the Smoluchowski equation \eqref{eq:smol}, the orientation dynamics is described by the rotational flux---the active particle exhibits rotary Brownian motion and is rotated by the fluid vorticity.

\subsection{Transport in the membrane}
\label{sec:membrane}
We treat the fluid transport in the membrane using a macroscopic approach similar to Darcy’s law; however, the porous region is ultimately modelled as a thin permeable interface. To this end, we first consider the membrane as having a network stress $\bsigma^\mathrm{net}$ and a fluid stress $\bsigma_f^m$. The defining characteristic of the semi-permeable membrane is that the fluid stress in the membrane balances the seepage velocity \citep{Durlofsky1987}:
\begin{equation}
    \label{eq:membrane-fluid-momentum}
    \bnabla\bcdot\bsigma_f^m - \eta_e\bR_m \bcdot\bu^s=\bm{0},
\end{equation}
or $\bu^s = \bK \bcdot\bnabla\bcdot\bsigma_f^m /\eta_e$, where $\bK = \bR_m^{-1}$ is the permeability tensor and $\bR_m$ is the membrane resistivity. The remaining network stress is responsible for maintaining the osmotic pressure difference across the membrane. That is, we have the force balance on the exterior and interior surfaces, respectively,
\begin{eqnarray}
    \label{eq:fluid-balance-membrane-interior}
    \bsigma_f^e \bcdot \bn &=& \bsigma_f^m \bcdot\bn,\quad \bx \in S_e\\
    \label{eq:fluid-balance-membrane-exterior}
    \bsigma_f^i \bcdot\bn &=& \bsigma_f^m\bcdot\bn,\quad \bx \in S_i.
\end{eqnarray}
Note, importantly, that at the interior surface, $\bsigma_f^i$ is the interior \emph{fluid} stress (it does not contain the osmotic pressure).

We model the membrane as a tangentially isotropic material with the permeability tensor
\begin{equation}
    \bK(\bn) = K_\perp \bn\bn + K_\parallel (\bI - \bn\bn),
\end{equation}
where $K_\perp$ is the normal permeability and $K_\parallel$ is the tangential one. For a thin membrane the gradient in equation \eqref{eq:membrane-fluid-momentum} can be approximated by a finite difference in the normal direction, which after applying the boundary conditions \eqref{eq:fluid-balance-membrane-interior} and \eqref{eq:fluid-balance-membrane-exterior}  leads to
\begin{eqnarray}
    \label{eq:seepage-constitutive}
   \bu^s(\bn) = \frac{\bK}{\eta_e \ell_m}\bcdot\left( \bsigma_f^e\big\rvert_{S_e} -\bsigma_f^i \big\rvert_{S_i}\right)\bcdot\bn.
\end{eqnarray}
Here,  $\ell_m$ is the thickness of the membrane and the thin membrane condition is $\ell_m \ll R$ with $R$ being the radius of the exterior surface. It is understood that in equation \eqref{eq:seepage-constitutive} $\bu^s$ is a function of the local outward normal vector $\bn$ (see figure \ref{fig:problem-schematic}). Equation \eqref{eq:seepage-constitutive} is a linear  relation that specifies how a seepage velocity is generated in response to a jump in the fluid stress across the membrane.

In the absence of deviatoric stress, equation \eqref{eq:seepage-constitutive} reduces to
\begin{equation}
    \bu^s = - \frac{K_\perp}{\eta_e\ell_m}\left( p_f^e\rvert_{S_e} -p_f^i \big\rvert_{S_i}\right)\bn,
\end{equation}
which is the more familiar Darcy's law in terms of the fluid pressure difference. In general, the normal flow is driven by the fluid pressure difference as well as the shear stress.

We remark that different boundary conditions across membranes and macroscopic transport equations exist in the literature. For example, an empirical boundary condition was proposed by \citet{beavers_joseph_1967} and later rationalized by \citet{Saffman1972}. This boundary condition was then generalized to a curved surface \citep{jones_1973}. Recently, using  multiscale   homogenization   and   matched   asymptotic   expansions between  the  near  membrane  and  the  far region, \citet{zampogna_gallaire_2020} developed a macroscopic condition to simulate  the  interaction  between  an  incompressible  fluid flow  and  a  permeable  thin membrane. For the purpose of the present work, equation \eqref{eq:seepage-constitutive} is sufficient.

Because the vesicle is rigid, the preservation of its volume dictates that
\begin{eqnarray}
 \int_{S_e} \bu^s \bcdot\bn dS =0.
\end{eqnarray}
Henceforth, for simplicity we shall assume that the membrane is not permeable in the tangential directions ($K_\parallel=0$), in which case the seepage velocity is normal to the vesicle surface.

In the above consideration, the vesicle membrane is treated as a rigid and thin porous region. To understand the material response of the vesicle, a proper treatment taking into consideration the constitutive law of the vesicle membrane is needed \citep{Lebedev2007,Vlahovska2007}. In particular, the bending elasticity and local incompressibility give rise to a surface force density in the membrane, which is balanced by the jump in the traction from the fluid inside and the fluid outside the vesicle membrane.  When such effects are included, the shape of the membrane is not known \textit{a priori} and must be determined as part of the solution. If the departure from the spherical shape is small, a perturbative approach can be adopted for both the membrane dynamics \citep{Lebedev2007,Vlahovska2007} and the fluid mechanics of a nearly spherical particle moving in a viscous fluid \citep{Brenner1964}.

\subsection{Dynamics of the vesicle}
The rigid-body translational and rotational velocities of the vesicle are determined by the force/torque-free conditions given by
\begin{eqnarray}
    \int _{S_e}\bsigma_f^e \bcdot\bn dS = \bm{0},\quad \int_{S_e}\bx \times \bsigma_f^e \bcdot\bn d S = \bm{0}.
\end{eqnarray}
We can relate the rigid-body velocities $\bU$ and $\bOmega$ to the seepage velocity $\bu^s$ at the exterior surface using the reciprocal theorem for Stokes flow \citep{masoud_stone_2019}. The formula for a general body shape is given in \citet{Elfring15}. For the case of a spherical particle, the rigid-body translational and rotational velocities are given by, respectively,
\begin{eqnarray}
    \bU = - \frac{1}{4 \pi R^2}\int_{S_e}\bu^s d S,\quad \bOmega = - \frac{3}{8\pi R^3} \int_{S_e} \bn\times \bu^s d S.
\end{eqnarray}
In the study of the rigid-body motion of micro-swimmers with prescribed kinematics (gaits) such as squirmers, the reciprocal theorem allows one to bypass the calculation of  the  unknown  flow  field,  provided  one  can  solve  the  resistance/mobility  problem for  the  swimmer  shape. For the problem considered here, the seepage velocity of the vesicle is not known \emph{a priori};  we need to determine the rigid-body motion, the exterior/interior flow fields and the distribution of ABPs simultaneously.

\subsection{Non-dimensional equations for a spherical vesicle}
For a spherical vesicle, the angular velocity vanishes ($\bOmega = \bm{0}$) and the torque balance is automatically satisfied. We define a characteristic swim speed $\tilde{U}_s$ such that
\begin{equation}
    U_s(\bx) = \tilde{U}_s \hat{U}_s(\bx).
\end{equation}
For a spatially homogeneous swim speed, $\hat{U}_s(\bx) =1$. The average density of ABPs inside the vesicle is $\overline{n} = N/V_i$, where $V_i = 4\pi (R-\ell_m)^3/3$ is the volume of the interior. We use this average density to scale the probability density such that
\begin{equation}
    \varPsi = \overline{n} g,
\end{equation}
where $g$ is the non-dimensional probability density. To render the governing equations non-dimensional, we scale pressures and stresses by $\overline{n}\ksts$, length by $R$ and fluid/vesicle velocities by $\overline{n}\ksts K_\perp/(\eta_e\ell_m)$. Recall that the activity $\ksts=\zeta\tilde{U}_s^2\tau_R/6$.

Using the characteristic swim speed, we define the swim P\'eclet number
\begin{equation}
   \Pes = \frac{\tilde{U}_s \tau_D}{R} = \frac{\tilde{U}_s R}{D_T}
\end{equation}
that compares the swim speed to the diffusive speed $R/\tau_D$, where $\tau_D = R^2/D_T$ is a diffusive timescale. Another dimensionless parameter for ABPs is defined as
\begin{equation}
    \gamma  = \sqrt{\frac{\tau_D}{\tau_R}} = \frac{R}{\delta},
\end{equation}
where $\delta = \sqrt{D_T \tau_R}$ is a microscopic length that quantifies the distance traveled by translational diffusion on the timescale of $\tau_R$. Alternate parameters including $\ell/\delta$ and $\ell/R$ are often used in the literature. These parameters are direct comparisons between different length scales. We note that they are related to $\Pes$ and $\gamma$ by $\Pes = (\ell/\delta)^2 (\ell/R)^{-1}$ and $\gamma = (\ell/R)^{-1}\ell/\delta$.

The non-dimensional exterior problem is given by
\begin{eqnarray}
    Da \nabla^2 \bu^e &=& \bnabla p_f^e,\label{eq:exterior-momentum}\\
    \bnabla\bcdot\bu^e&=&0,\label{eq:exterior-continuity}\\
    \bu^e &\to& \bm{0}\quad\mbox{and} \quad p_f^e \to 0\quad\mbox{as}\quad r \to \infty,\\
    \bu^e &=& \bU+ \bu^s\quad\mbox{at}\quad r = 1.\label{eq:exterior-bc-nondim}
\end{eqnarray}
where
\begin{equation}
    Da = \frac{K_\perp}{R \ell_m},
\end{equation}
is a Darcy number that compares the permeability of the membrane to its characteristic cross-sectional area.

In the interior, the rigid-body translation $\bU$ has no effect on the fluid dynamics and we only need to consider the deviation $\bu^\prime = \bu^i - \bU$. Thus, the non-dimensional flow problem in the interior is governed by
\begin{eqnarray}
    \beta Da \nabla^2 \bu^\prime &=& \bnabla P,\label{eq:interior-momentum}\\
    \bnabla\bcdot\bu^\prime&=&0,\label{eq:interior-continuity}\\
    |\bu^\prime|, P  &<& \infty \quad \mbox{at}\quad  r=0,\\
    \bu^\prime &=&  \bu^s \quad \mbox{at}\quad  r = \Delta.\label{eq:interior-bc-nodim}
\end{eqnarray}
Here,
\begin{equation}
    \beta = \frac{\eta_i}{\eta_e}
\end{equation}
is the interior-to-exterior viscosity ratio and
\begin{equation}
    \Delta = \frac{R-\ell_m}{R}
\end{equation}
is the radius ratio between the interior and the exterior surfaces of the membrane. For a thin membrane, $\ell_m/R\ll 1$, $\Delta$ is $O(1)$. The non-dimensional total pressure is given by
\begin{equation}
\label{eq:non-dimensional-pressure-interior}
    P = p_f^i + \frac{k_BT}{k_sT_s} n= p_f^i + \frac{6\gamma^2}{\Pes^2} n,
\end{equation}
where we have used the relation $k_BT/k_sT_s = D_T / (\overline{U}_0^2\tau_R/6) = 6 \gamma^2/\Pes^2$.

The non-dimensional deviatoric stress tensors in the exterior and interior are, respectively,
\begin{eqnarray}
\label{eq:deviatoric-non-dimensional}
    \btau^e =Da \left[ \bnabla \bu^e + \left(\bnabla \bu^e\right)^T \right],\quad \btau^i = \beta Da \left[ \bnabla \bu^\prime + \left(\bnabla \bu^\prime \right)^T \right].
\end{eqnarray}
The seepage velocity is given by
\begin{equation}
\label{eq:non-dimensional-seepage}
    \bu^s = \bn\bn\bcdot\left( \bsigma_f^e\big\rvert_{S_e} -\bsigma_f^i \big\rvert_{S_i}\right)\bcdot\bn,
\end{equation}
where $\bsigma_f^e = -p_f^e \bI + \btau^e$ and $\bsigma_f^i = -p_f^i \bI + \btau^i$. The volume conservation of the vesicle is
\begin{equation}
\label{eq:volume-conservation}
    \int_{S_e} \bu^s \bcdot \bn = 0.
\end{equation}
The rigid-body translational velocity of the vesicle is then
\begin{equation}
\label{eq:reciprocal-dimensionless}
    \bU = -\frac{1}{4\pi}\int_{S^2} \bu^s d\Omega.
\end{equation}

The non-dimensional Smoluchowski equation, its fluxes, boundary condition and particle conservation are, respectively,
\begin{eqnarray}
\label{eq:dimensionless-smol}
    &&\bnabla\bcdot \bj_T + \bnabla_R\bcdot\bj_R=0,\\
    \label{eq:dimensionless-jT}
    &&\bj_T = \alpha Da \bu^\prime g + \Pes \hat{U}_s(\bx) \bq g - \bnabla g,\\
    \label{eq:dimensionless-jR}
    &&\bj_R = \frac{1}{2}\alpha Da \bomega^\prime g - \gamma^2 \bnabla_R g,\\
    &&\bn\bcdot\bj_T = 0\quad\mbox{at}\quad r=\Delta,\\
    &&\int g d \bq d\bx = \frac{4\pi}{3}\Delta^3, \label{eq:non-dimensional-conservation}
\end{eqnarray}
where we have introduced three non-dimensional parameters $\alpha, \Pes$ and $\gamma$. The first parameter is a reduced osmotic pressure and given by
\begin{equation}
    \alpha = \frac{\nb \ksts \tau_D}{\eta_e}.
\end{equation}
Physically, this is a comparison between the active driving pressure ($\overline{n}\ksts$) and a viscous resistive `pressure' ($\eta_e/\tau_D$) on the timescale $\tau_D$.

In the equations above, variables $\{\bu^e, p_f^e, \bx, r, \bU, \bu^s, P, \bu^\prime\}$ and gradient operators are non-dimensional even though the same symbols as their dimensional counterparts are used. This is to avoid inconvenience in notation and henceforth we shall work with non-dimensional quantities unless otherwise noted.

\begin{table}
  \begin{center}
\def~{\hphantom{0}}
  \begin{tabular}{cll}
     Non-dimensional parameter  & Mathematical definition &  Physical description  \\[3pt]
      \hline
       $\alpha$  &$\nb\ksts \tau_D/\eta_e$   & Reduced osmotic pressure\\
       $\beta$ & $\eta_i/\eta_e$   & Viscosity ratio\\
       $\gamma$ & $ R /\delta $ &  Comparison of $R$ and $\delta$ \\
       $Da$ & $K_\perp/(R\ell_m)$ & Darcy number\\
       $\Pes$ &  $\tilde{U}_s\tau_D/R$ & Swim P\'eclet number\\
       $\Delta$ & $(R-\ell_m)/R$ & Radius ratio\\
   \end{tabular}
  \caption{Independent non-dimensional parameters.}
  \label{tab:parameters}
  \end{center}
\end{table}

It is convenient to consider the orientational moments of the probability density function. The zeroth order moment, or the number density is given by
\begin{equation}
    n(\bx) = \int_{S^2}g d\bq,
\end{equation}
where $S^2$ is the surface of the unit sphere in $\mathbb{R}^3$, which represents all possible orientations that $\bq$ takes. Integrating the Smoluchowski equation over all orientations, we obtain a conservation equation for the number density
\begin{subequations}
\begin{equation}\label{eq:number-equation}
    \bnabla\bcdot\bj_{n}=0,
\end{equation}
\begin{equation}
    \bj_n = \alpha Da \bu^\prime n + \Pes \hat{U}_s(\bx) \boldm -  \bnabla n.
\end{equation}
\end{subequations}
This equation is coupled to the first moment, or polar order,
\begin{equation}
    \boldm(\bx) = \int_{S^2} \bq g d\bq.
\end{equation}
The no-flux condition \eqref{eq:noflux} becomes $\bn\bcdot\bj_n =0$ for $\bx\in S_i$. Multiplying the Smoluchowski equation by $\bq$ and integrating over $S^2$, we obtain a governing equation for the polar order,
\begin{subequations}
\begin{equation}\label{eq:polar-equation}
    \bnabla\bcdot\bj_{m} - \frac{1}{2}\alpha Da \bomega^i\times \boldm + 2 \gamma^2 \boldm=\bm{0},
\end{equation}
\begin{equation}
    \bj_{m} = \alpha Da \bu^\prime \boldm + \Pes \hat{U}_s(\bx)\left(  \bQ  + \frac{1}{3}n \bI\right)  -\bnabla\boldm
\end{equation}
\end{subequations}

where
\begin{equation}
    \bQ = \int_{S^2} \left(\bq\bq-\frac{1}{3}\bI\right)g d\bq,
\end{equation}
 is the trace-free nematic order tensor and $\bI$ is the identity tensor of rank two. The no-flux condition at the interior surface for the polar order becomes $\bn\bcdot\bj_m = \bm{0}$.  Different from the conservation of the total number of ABPs, the polar order is not conserved as indicated by the presence of the sink term $2\gamma^2\boldm$ in equation \eqref{eq:polar-equation} even in the absence of flow. This sink term describes the randomization, due to rotary diffusion, of any polar order.

As can be inferred from the above discussion, there is an infinite hierarchical structure to the moment equations. To truncate this infinite set of equations, a closure model such as $\bQ =\bm{0}$ is often considered in the literature~\citep{Saintillan2015,yan_brady_2015}. A closure leads to a set of closed equations that can be solved as an approximation to the Smoluchowski equation. We note that a closure approximation is often not uniformly accurate across different regimes of physical parameters or different spatial/time domains and care must be taken when interpreting results obtained from such methods~\citep{Dulaney2020,Burkholder2020,Peng2020}. A systematic approach to derive low-order closure models that are able to approximate the full solution of the Smoluchowski equation is still lacking.

In the context of active nematic (apolar) suspensions, the Bingham closure \citep{Chaubal1998} has been shown to agree well with the full kinetic theory and recently a numerical scheme has been developed to efficiently evaluate the Bingham closure \citep{WEADY2022110937}. With this closure, simulations with high spatial resolution are performed for active nematics. As note by \citet{WEADY2022110937}, their closure is formulated for apolar suspensions and the generalization to polar active matter remains (e.g., ABPs) to be considered. Furthermore, a comparison of the accuracy of different closure models for ABPs is largely unexplored.

The mechanism for an induced concentration gradient from a prescribed activity gradient in the absence of flow has been studied in previous works~\citep{Schnitzer1993,Tailleur2008,Row2020}. To illustrate this mechanism and motivate later discussions, we summarize the simple one-dimensional (1D) result here. In the absence of external linear or angular velocities, such as due to flow or orienting field, the governing equation in 1D for highly active ABPs is $\bnabla\bcdot(\hat{U}_s\boldm)=0$, where the diffusive term is neglected. The solution in 1D is simply $\boldm=0$. Then, equation \eqref{eq:polar-equation} reduces to $n \hat{U}_s = const$. Further, \cite{Row2020} showed that this spatial variation of activity and concentration can drive a reverse osmotic flow, i.e. fluid flow from regions of high concentration to low. In this work, we exploit this spatial variation to propel a vesicle that is able to maintain an activity gradient in the swim speed of ABPs confined inside.

\section{Vesicle motion in the limit of weak interior flow}
\label{sec:weak-interior-flow}
In many situations, the advection due to the interior fluid flow is much weaker compared to the self-propulsion of the ABPs or its active swim diffusion (small P\'eclet number), and we may neglect the effect of the fluid velocity disturbance on the distribution of ABPs.

\subsection{Governing equations}
\label{sec:weak-permeability}
The behavior of the system in this small-P\'eclet limit can be systematically derived by considering a weakly permeable membrane, $Da \ll 1$.

If the vesicle is non-permeable ($Da = 0$), no external or internal flows can be generated, and the vesicle remains stationary despite the nonuniform density distribution and accumulation of the ABPs at the boundary.  Due to the scaling of the dimensional velocities by the permeability, the leading order non-dimensional velocities are $O(1)$ as $Da \to 0$. To study the motion of the vesicle in the $Da \ll 1$ limit, we pose regular expansions for all fields:
\begin{eqnarray}
    \bu^e &=& \bu_0^e + Da \bu_1^e +\cdot\cdot\cdot, \\
    p_f^e &=& p_{f,0}^e +Da p_{f,1}^e  +\cdot\cdot\cdot,\\
    \bu^\prime &=& \bu_0^\prime + Da \bu_1^\prime +\cdot\cdot\cdot,\\
    P &=& P_0 +Da P_1  +\cdot\cdot\cdot,\\
    g&=& g_0 +Da g_1 +\cdot\cdot\cdot.
\end{eqnarray}
The dimensionless number density is given by $n=\int gd\bq = n_0 +Da n_1+\cdot\cdot\cdot.$
Similarly, the expansions for the translational and the seepage velocities are, respectively,
\begin{eqnarray}
    \bU &=& \bU_0 + Da \bU_1  +\cdot\cdot\cdot,\\
    \bu^s&=& \bu_0^s +Da \bu_1^s +\cdot\cdot\cdot.
\end{eqnarray}
From equation \eqref{eq:deviatoric-non-dimensional}, we know that the leading order deviatoric stresses are $O(Da)$, which does not contribute to the $O(1)$ seepage velocity. As a result, the seepage velocity at leading order is driven by the fluid pressure difference across the membrane,
\begin{eqnarray}
\label{eq:seepage-Da0}
    \bu_0^s = \left(p_f^i \big\rvert_{S_i} - p_f^e \big\rvert_{S_e}  \right)\bn.
\end{eqnarray}

Inserting these expansions into the exterior Stokes equations \eqref{eq:exterior-momentum} and \eqref{eq:exterior-continuity} gives to leading order
\begin{eqnarray}
\label{eq:exterior-Da0}
    \bnabla p_{f,0}^e = 0, \quad \bnabla\bcdot\bu_0^e=0.
\end{eqnarray}
The kinematic boundary condition at the exterior surface is $\bu_0^e(r=1) = \bU_0 + \bu_0^s$. Due to the linearity of Stokes flow, we only need to solve equation \eqref{eq:exterior-Da0} using the seepage velocity condition [$\bu_0^e(r=1) = \bu_0^s$]; the rigid body translation is determined from the reciprocal theorem given by equation \eqref{eq:reciprocal-dimensionless}. Because $\bu_0^s$ is in the radial direction, the exterior flow is radial and given by
\begin{equation}
    p_{f,0}^e = 0,\quad \bu_0^e = \frac{\bu_0^s}{r^2}.
\end{equation}
Similarly, the leading order equation governing the interior flow is given by
\begin{eqnarray}
    \label{eq:interior-Da0}
    \bnabla P_{0} = 0, \quad \bnabla\bcdot\bu_0^\prime=0.
\end{eqnarray}
At the interior surface, the flow field satisfies the condition $\bu^\prime(r=\Delta) = \bu^s$. We note that the interior flow field is not analytically tractable but it is not required in order to determine the vesicle motion. The total pressure at leading order is a constant, consisting of spatially varying fluid pressure and osmotic pressure,
\begin{equation}
\label{eq:P-Da-0}
    p_{f,0}^i +  6\gamma^2 n_0/\Pes^2=P_0 =const.
\end{equation}
 Inserting the expansions into the Smoluchowski equation \eqref{eq:dimensionless-smol}--\eqref{eq:non-dimensional-conservation}, we obtain at leading order
\begin{eqnarray}
\label{eq:smol-no-flow}
 &&\bnabla\bcdot \left(\Pes \hat{U}_s(\bx) \bq g_0 - \bnabla g_0  \right)- \gamma^2 \bnabla_R^2 g_0=0,\\
 \label{eq:smol-bc-no-flow}
 &&\bn\bcdot\left(\Pes \hat{U}_s(\bx) \bq g_0 - \bnabla g_0  \right)=0\quad\mbox{at}\quad r=\Delta,\\
  \label{eq:smol-conservation-no-flow}
 &&\int g_0 d \bq d\bx = \frac{4\pi}{3}\Delta^3.
\end{eqnarray}
Using equations \eqref{eq:reciprocal-dimensionless}, \eqref{eq:seepage-Da0} and \eqref{eq:P-Da-0}, we obtain
\begin{equation}
\label{eq:U-Da-0}
    \bU_0 = \frac{3\gamma^2}{2\pi \Pes^2} \int_{S^2} n_0(r=\Delta) \bn d \Omega.
\end{equation}
It is more intuitive to examine the above expression in its dimensional form
\begin{equation}
\label{eq:U-Da-0-dimensional}
    \bU_0 = \frac{1}{4\pi} \frac{K_\perp}{\eta_e \ell_m} \int_{S^2} \Pi^\mathrm{osmo}_0 \bn d \Omega,
\end{equation}
where $\Pi_0^\mathrm{osmo} = n^w k_BT$ is the dimensional osmotic pressure of ABPs in the absence of flow.

To sum up, one needs to solve equations \eqref{eq:smol-no-flow}-\eqref{eq:smol-conservation-no-flow} to obtain the density distribution of ABPs in the absence of flow, and then using equation \eqref{eq:U-Da-0} to calculate the vesicle motion. In the remainder of section \ref{sec:weak-interior-flow}, the subscript `0' (e.g., $g_0$, $\bU_0$) will be dropped for notational convenience.

In general, one can represent the number density distribution at the spherical interior wall by the complete spherical harmonic expansion
\begin{equation}
    n_0(\Delta, \theta, \phi) = \sum_{l=0}^\infty\sum_{m=-l}^{m=l}C_{l,m} Y_l^m(\theta, \phi),
\end{equation}
where $Y_l^m = \sqrt{(2l+1)(l-m)!/[4\pi(l+m)!]}P_l^m(\cos\theta)\exp(i m \phi)$ and $P_l^m$ is the associated Legendre polynomial of degree $l$ and order $m$. Using equation \eqref{eq:U-Da-0}, a direct integration shows that only the $l=1$ modes contribute to the  translational velocity of the vesicle. This is similar to the tangential spherical squirmer model in which only the ``$B_1$'' mode---the coefficient of $P_1^1(\cos\theta)$--- contributes to the velocity of the squirmer.

\subsection{High activity}
\label{sec:high-activity-Da}
We now explore the limit of high activity, $\ksts/\kt = \tilde{D}^\mathrm{swim}/D_T = \ell^2/(6\delta^2)\gg 1$, which is often observed in active matter systems \citep{takatori2016acoustic}.  Equivalently, we define $\epsilon  = 1/\gamma^2$ (Note that $\Pes = \gamma^2 \ell/R$) and consider the limit $\epsilon \to 0$. Expanding the probability density function $g = g^{(0)}+\epsilon g^{(1)}+\cdot\cdot\cdot$, we obtain at leading order
\begin{equation}
    \label{eq:bulk-high-activity}
    \frac{\ell}{R} \bnabla\bcdot \left[  \hat{U}_s \bq g^{(0)}\right] - \frac{1}{\hat{\tau}_R}\nabla^2_R g^{(0)}=0,
\end{equation}
where we have included the spatial variation of $\tau_R(\bx)$ and defined $\tau_R = \tilde{\tau}_R\hat{\tau}_R$ similar to the case of spatially varying swim speed. Integrating over the orientation space leads to an equation for the polar order
\begin{equation}
    \bnabla\bcdot\left( \hat{U}_s \boldm^{(0)} \right)=0.
\end{equation}
Equation \eqref{eq:bulk-high-activity} is incompatible with the no-flux boundary condition and thus is only valid in the bulk of the interior. At the interior membrane surface, the swimming flux is balanced by the diffusive flux, which implies the existence of an accumulation boundary layer of thickness $O(\epsilon)$. In this high activity limit, the number of particles in the boundary layer is still finite, which suggests that the probability density is $O(1/\epsilon)$ as $\epsilon\to 0$. Therefore, the probability density in the boundary layer admits an expansion of the form $g(y, \theta, \phi, \bq) = g^{(-1)}/\epsilon + g^{(0)} +\cdot\cdot\cdot$. Defining a stretched boundary-layer coordinate in the radial direction $y = (\Delta -r)/\epsilon$, the Smoluchowski equation to leading order is
\begin{eqnarray}
        \label{eq:eq-high-activity-Da}
       \frac{\ell}{R} \hat{U}_s\big\rvert_{S_i} \bq\bcdot\be_r \frac{\partial g^{(-1)}}{\partial y} + \frac{\partial^2 g^{(-1)}}{\partial y^2}&=&0,\\
       \label{eq:bc-high-activity}
       \frac{\ell}{R} \hat{U}_s\big\rvert_{S_i} \bq\bcdot\be_rg^{(-1)} + \frac{\partial g^{(-1)}}{\partial y}&=&0\quad\mbox{at}\quad y=0,\\
       g^{(-1)} \to 0 \quad\mbox{as}\quad y\to +\infty.
\end{eqnarray}
Here, the Taylor expansion $\hat{U}_s(r, \theta, \phi) = \hat{U}_s\big\rvert_{S_i} -\epsilon y \frac{d\hat{U}_s}{dr}\big\rvert_{S_i}+\cdot\cdot\cdot$ is used. The solution is readily obtained
\begin{equation}
\label{eq:g-high-activity}
    g^{(-1)} =
    \begin{cases}
    A_1(\theta, \phi,\bq) \exp\left( -\frac{\ell}{R} \hat{U}_s\big\rvert_{S_i} \bq\bcdot\be_r y\right), & \bq\bcdot\be_r >0,\\
     0, & \mathrm{otherwise}.
    \end{cases}
\end{equation}
This singular accumulation only occurs for particles with orientation pointing towards the wall ($\bq\bcdot\be_r >0$) because otherwise they would swim away. In equation \eqref{eq:g-high-activity}, $A_1$ is an unknown function that can only be determined from the next-order solution. The boundary-layer solution $g^{(0)}(y, \theta,\phi,\bq)$ in the limit $y\to\infty$ needs to be matched with the solution in the bulk as $r\to \Delta$.

At the interior surface of the vesicle ($y=0$), the leading-order density is large and given by $\gamma^2 \int_{\bq\bcdot \be_r >0} A_1 d\bq$. Just outside the boundary layer (i.e., $y\to \infty$), the density is $O(1)$ as $\gamma^2 \to \infty$. This boundary-layer structure allows us to relate the osmotic pressure at the interior surface of the vesicle to the swim pressure outside the boundary layer. To this end, we consider the ratio $n^w \kt/\left(n^0\ksts\right)$, where all quantities are dimensional. The density at the wall $n^w$ and the density outside the boundary layer $n^0$ are defined locally along the interior surface and are functions of the local surface normal vector $\bn$. From the above analysis, we have
\begin{equation}
    \label{eq:nwkt-n0ksts}
    \frac{n^w\kt}{n^0\ksts} = \frac{\gamma^2 \int_{\bq\bcdot \be_r >0} A_1 d\bq }{\int g^{(0)}(y\to \infty, \theta,\phi,\bq) d\bq} \frac{\kt}{\ksts} = f(\ell/R, \Delta),
\end{equation}
where $\gamma^2\kt/\ksts=6R^2/\ell^2$ is not a function of the thermal diffusivity $D_T$ (or $\ell/\delta$). Because in general $A_1$ is not analytically tractable,  the  factor $f(\ell/R, \Delta)$  in the preceding equation cannot be explicitly obtained. Nevertheless, equation \eqref{eq:nwkt-n0ksts} reveals the important fact that at high activity
\begin{equation}
    \label{eq:Pi-osmo-pi-swim}
    \Pi^\mathrm{osmo} = n^w\kt = \Pi^\mathrm{swim}_0 f(\ell/R, \Delta),
\end{equation}
 where $\Pi^\mathrm{swim}_0 = n^0\ksts$. In other words, the osmotic pressure at the wall is equal to the swim pressure in the bulk of the interior just outside the boundary layer but modified by a scale factor that is a function of $\ell/R$ and $\Delta$. We emphasize that in equation \eqref{eq:Pi-osmo-pi-swim}, all quantities are defined locally along the interior surface of the vesicle.
This is a generalization of the  result of \citet{yan_brady_2015} for ABPs outside an infinite planar wall, where $n^w\kt = n^0\ksts$ in the limit $\gamma^2 \to \infty$ because of the absence of curvature of the geometry.

\begin{figure}
    \centering
    \includegraphics[width=0.7\textwidth]{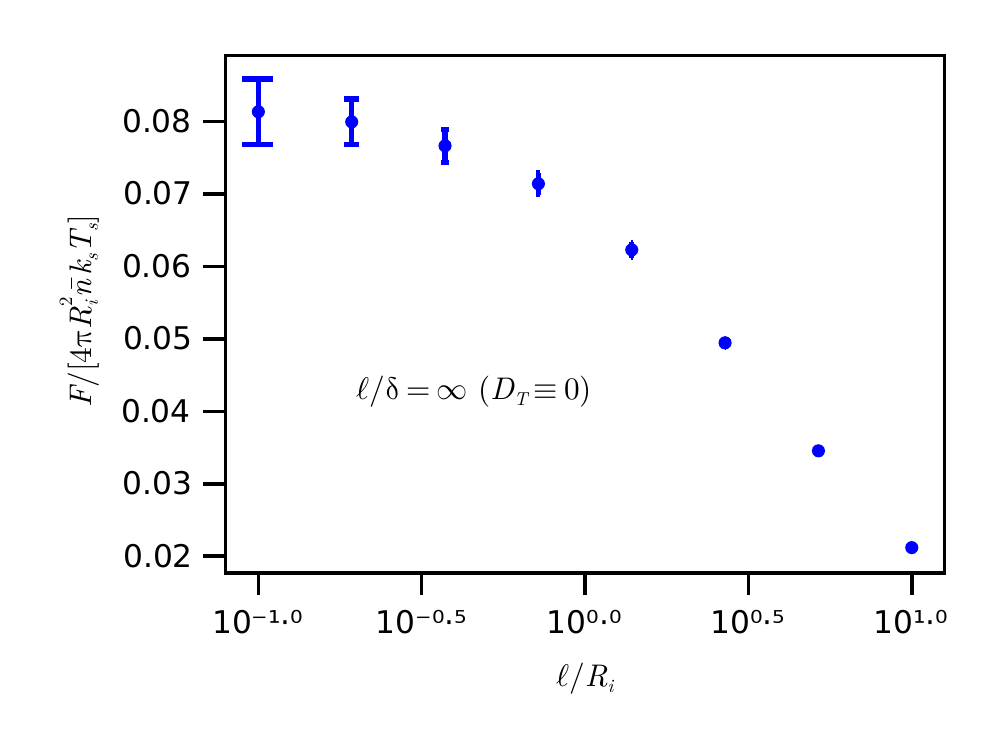}
    \caption{\label{fig:step-speed}The magnitude of the dimensionless net force on the interior vesicle surface $\bF^w/(4\pi R_i^2\overline{n}\ksts)$  as a function of $\ell/R_i$ for ABPs with spatially varying swim speed. The speed profile is a step function where the swim speed in one of the hemisphere is half of that in the other. The reorientation time $\tau_R$ is a constant. The net force points towards the side with a higher swim speed.}
\end{figure}

Equation \eqref{eq:Pi-osmo-pi-swim} allows us to obtain the dimensional speed of the vesicle:
\begin{equation}
    \label{eq:speed-high-activity-main}
    \bU = \frac{1}{4\pi} \frac{K_\perp}{\eta_e\ell_m}  \int_{S^2} \Pi_0^\mathrm{swim} f(\ell/R, \Delta) \bn d \Omega.
\end{equation}
We note that this relation holds for ABPs with spatially varying swim speed or reorientation time.

To understand the dependence of the motion of the vesicle on $\ell/R$, we approach the problem from a micromechanical perspective using Brownian dynamics simulations that resolve the Langevin equations of motion governing the stochastic dynamics of an ABP in its physical and orientation space. The details of the simulation method is given in section \ref{sec:BD}. The ABPs are treated as point particles and their hard-particle interaction with the vesicle interior boundary is implemented using the potential-free algorithm \citep{HEYES19931}. In this approach, the force exerted on the wall due to the collision with ABPs is readily obtained. Consider a simulation of $N$ ABPs that only interact with the boundary independently but not among themselves. After a time step $\Delta t$, some particles might have moved outside the interior wall. For particle $i$ that is now outside, we add a displacement $\Delta \bx_i$ to the particle such that after the move the particle is at contact with the boundary. The total force exerted on the wall is then $\bF^w=-\zeta \sum_{i \in \mathcal{I} }\Delta \bx_i/\Delta t$ where $\mathcal{I}$ is the set of all particles that are outside the boundary before the hard-sphere move. As seen in equation \eqref{eq:U-Da-0-dimensional}, the net speed of the vesicle is proportional to the net force $\bF^w$.

In figure \ref{fig:step-speed}, we show the dimensionless net force exerted on the interior vesicle surface by the ABPs, $\bF^w/(4\pi R_i^2\overline{n}\ksts)$,  as a function of $\ell/R_i$ for ABPs with no $D_T$ (infinitely active, $\ell/\delta=\infty$) and a spatially-varying swim speed. The swim speed profile is a step function given by
\begin{equation}
    \label{eq:Us-step}
    \hat{U}_s = \begin{cases}
        1 & x < 0,\\
        1/2 & x >0.
    \end{cases}
\end{equation}
The net force points to the side with a larger swim speed and only the force magnitude is shown in figure  \ref{fig:step-speed}. As $\ell/R_i$ increases, the net force decreases. For large $\ell/R_i$, the ABPs spend most of their time pushing against and sliding along the interior vesicle surface until rotary Brownian motion reorients them towards the bulk of the interior. In this limit, the number of particles pushing against the interior surface on the side of slow speed is comparable to the side of high speed.

As discussed earlier, in 1D the relation $n U_s=const$ holds for ABPs with spatially varying properties. In the interior of a vesicle, this relation is still useful for the qualitative understanding of the distribution of ABPs and the motion of the vesicle. Taking the step-function given by equation \eqref{eq:Us-step} as an example,  $n^0 U_s=const$ means that in the bulk of the interior the density on the right side ($x>0$) is higher than that on the left ($x<0$), $n^0_R > n^0_L$. Because $n^w \sim n^0 \ksts/\kt \sim n^0 U_s \zeta \ell/\kt$ and $n^0U_s=const$, we have $n^w \sim U_s$ for ABPs with constant $\tau_R$. Therefore, the density at the interior vesicle surface on the right side is \emph{lower} than that on the left ($n^w_R < n^w_L$), which is \emph{opposite} to the behavior of the bulk density. Because only the ABPs at the interior surface contribute to the net force, and they can only push against the boundary, this leads to the fact that the net force is in the negative $x$ direction (to the left). If one only had observations of the number density in the bulk, one would conclude that the vesicle moves in the direction of a lower concentration---a `reverse' osmotic propulsion [cf. equation \eqref{eq:intro-linear-grad}].

\begin{figure}
    \centering
    \includegraphics[scale=1.0]{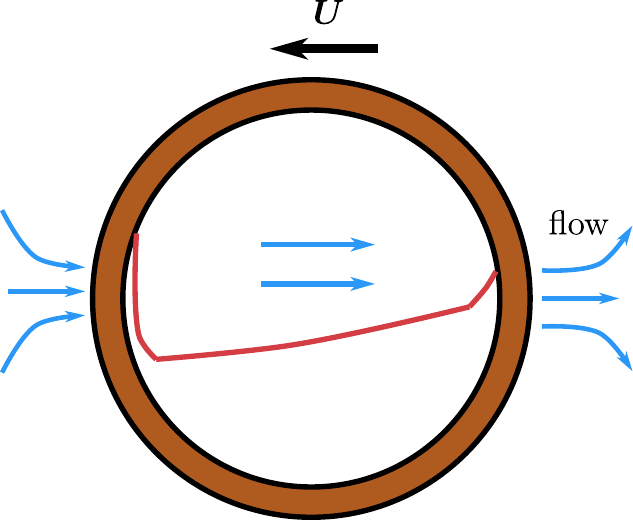}
    \caption{\label{fig:boundary-layer-sketch}Schematic of the number density profile (red) and the flow direction (blue) in the high activity limit for a swim-speed profile that decreases from the left to the right. A weak density gradient is present in the bulk of the interior due to the variation of the swim speed. Two accumulation boundary layers are established at the left and right sides of the interior wall, with the density at the wall on the left larger than that on the right. The vesicle-ABPs system as a whole moves by way of jet propulsion. }
\end{figure}

The number density profile in the bulk and the boundary layer is sketched in figure \ref{fig:boundary-layer-sketch} (red line) for a general swim-speed profile that  decreases from the left to the right. The variation of the swim speed leads to a gradient in the number density in the bulk of the interior. Two thin accumulation boundary layers are established at the left and right sides of the interior vesicle surface. Because the density at the wall on the right is smaller than that on the left, $n^w_R < n^w_L$, the dimensional version of equation \eqref{eq:P-Da-0} then leads to a larger fluid pressure on the low density side (right), $p_{f,R}^i > p_{f,L}^i$. Since the fluid pressure in the exterior is homogeneous, the fluid is pushed out of the vesicle from the right and drawn in from the left by conservation of mass. For the vesicle-ABPs system as a whole, it effectively moves by way of jet propulsion. This kind of noninertial jet propulsion has been proposed and studied in detail by \citet{Spagnolie10} as an alternate mechanism for the locomotion of microswimmers. In their paper, the jetting velocity distribution of a microswimmer ($\bu^s$) is prescribed, and then the swim speed is determined from the reciprocal theorem.

Using the approximation $n^0 U_s=const$ and the relation $\Pi_0^\mathrm{swim} = n^0\ksts =n^0 U_s\zeta \ell/6$, we see that it is the variation of run length $\ell(\bx)$ that is responsible for the net force on the vesicle interior surface and ultimately the vesicle motion. Using equation \eqref{eq:speed-high-activity-main}, a Taylor series expansion about the center of the vesicle leads to the scaling relation $\bU \sim K_\perp R \zeta n^0 U_s \bnabla \ell/(\eta_e\ell_m) $, where $\nabla\ell$ is the gradient of the run length at the center of the vesicle.

\subsection{A large vesicle}
\label{sec:large-vesicle}
When the vesicle is large, the confinement is weak, $\ell/R \ll 1$, ABPs exhibit a thin accumulation boundary layer at the wall and a uniform distribution in the bulk of the interior to leading order. To study this large-vesicle limit of $\ell/R \ll 1$, we first write equation \eqref{eq:smol-no-flow} equivalently as
\begin{equation}
    \bnabla\cdot\left[ \frac{\ell}{R} \hat{U}_s(\bx)\bq g - \left(\frac{\ell}{R}\right)^2 \left(\frac{\ell}{\delta}\right)^{-2}  \bnabla g \right] - \nabla_R^2 g=0.
\end{equation}
In this section, we use the definition $\epsilon = \ell/R$ and consider the limit as $\epsilon\to 0$. In the bulk of the interior, we have the expansion $g = g^{(0)} + \epsilon g^{(1)}+\cdot\cdot\cdot$ and the leading order equation $\nabla_R^2 g^{(0)}=0$. The solution in the bulk is then $g^{(0)}(\bx, \bq) = n^{(0)}(\bx)/(4\pi)$.  The boundary-layer thickness is determined by a balance between the swimming and the diffusive fluxes, which leads to the leading-order equation
\begin{eqnarray}
    \label{eq:large-vesicle-BL}
    - \frac{\partial}{\partial \rho} \left( \frac{\ell}{\delta} \hat{U}_s\big\rvert_{S^i}\bq\bcdot\be_r g^{(0)} + \frac{\partial }{\partial \rho}g^{(0)}\right) - \nabla_R^2 g^{(0)}&=&0,\\
    \frac{\ell}{\delta} \hat{U}_s\big\rvert_{S^i} \bq\bcdot\be_r g^{(0)} + \frac{\partial }{\partial \rho}g^{(0)} &=&0\quad\mbox{at}\quad \rho=0.
\end{eqnarray}
Here, we have used the stretched coordinate $\rho = (\Delta-r)/\epsilon$. Since $\ell \ll R$, curvature of the domain has no effect at $O(1)$ and  the boundary-layer equation is similar to that in a planar domain. The $O(1)$ probability density in the boundary layer does not contribute to the $O(1)$ conservation because the boundary layer thickness is $O(\epsilon)$. This means that the total conservation is given by the density outside the boundary layer alone, $\int n^{(0)}(\bx) d \bx = 4\pi\Delta^3/3$. In the absence of curvature terms, just like the problem of ABPs on one side of an infinite planar wall \citep{yan_brady_2015}, the number density at the interior wall of the vesicle at $O(1)$ can be determined analytically; the result is given by
\begin{equation}
   \frac{n^w}{n^0  } = 1 + \frac{1}{6}\left( \frac{\ell}{\delta}\right)^2 \hat{U}^2_0\big\rvert_{S^i}.
\end{equation}

In dimensional terms, this means that the osmotic pressure at the wall $\Pi_0^\mathrm{osmo}=n^w\kt = n^0 \kt + n^0  \ksts \hat{U}_s^2$ where $n^0 $ is the density outside the boundary layer. To determine $n^0 $, one needs to solve equation \eqref{eq:large-vesicle-BL} and then match the boundary-layer solution to that in the bulk.

The dimensional translational velocity in the large-vesicle limit is written as
\begin{equation}
    \label{eq:speed-large-vesicle}
    \bU =  \frac{1}{4\pi} \frac{K_\perp}{\eta_e\ell_m}\int_{S^2} \left[n^0  \kt + n^0 \ksts \hat{U}_s^2\big\rvert_{S^i} \right]\bn d \Omega.
\end{equation}

For a large vesicle, the accumulation boundary layer has a similar structure to that obtained in the high-activity limit. Even for weakly active ABPs, this accumulation boundary layer exists so long as $\ell/R \ll 1$. As expected, equation \eqref{eq:speed-large-vesicle} reduces to a form of \eqref{eq:speed-high-activity-main} if the activity is high.

\subsection{Vesicle motion due to an external orienting field}
\label{sec:external-field}
Another way to achieve motion is to apply an external orienting field, which affects the orientational dynamics but not the swim speed of the ABPs.  \citet{TakatoriSM} showed that net directed motion of ABPs in free space can be achieved due to the fact that the external field can orient particles to move in the same direction. Instead of having ABPs with spatially varying swim speed, we consider the same orienting field as in \citet{TakatoriSM} but now with ABPs confined inside the vesicle. The only change to the orientational dynamics is that the orienting field exerts an external torque that depends on the orientation of the particle relative to the field direction; the dimensional rotary flux now becomes $\bj_R = \Omega_c \bq\times \hat{\bH} g - D_R \bnabla_R  g$, where $\Omega_c$ characterizes the rate of reorientation due to the field and $\hat{\bH}$ is the direction of the field. When an ABP is aligned with the field direction ($\bq \parallel \hat{\bH} $), the external torque vanishes. The Smoluchowski equation \eqref{eq:smol-no-flow} for ABPs with constant properties in the presence of an orienting field is then
\begin{equation}
    \label{eq:smol-field}
    \bnabla\bcdot \left(\Pes \bq g - \bnabla g  \right)+\gamma^2 \bnabla_R\bcdot\left(\chi_R \bq \times \hat{\bH} g - \bnabla_R g\right) =0,
\end{equation}
while the no-flux boundary condition \eqref{eq:smol-bc-no-flow} and the total conservation \eqref{eq:smol-conservation-no-flow} remain unchanged. Here, we have defined the Langevin parameter, $\chi_R=\Omega_c\tau_R$, which measures the strength of the orienting field compared to rotary diffusion.

In the high-activity limit, an accumulation boundary layer is established at the interior wall. The boundary-layer structure is identical to that obtained for ABPs with spatially varying swim speed. At leading-order, the probability density in the bulk of the interior is governed by
\begin{equation}
    \frac{\ell}{R} \bq\bcdot\bnabla  g^{(0)} +\bnabla\bcdot\left( \chi_R \bq \times \hat{\bH}  g^{(0)} - \bnabla_R g^{(0)} \right) =0.
\end{equation}
Compared to  \eqref{eq:bulk-high-activity} for spatial variation, the preceding equation has a constant swim speed and the orientational dynamics is affected by the orienting field. In the boundary layer, the leading-order equation is identical to \eqref{eq:eq-high-activity-Da} and the density at the wall is large.

\begin{figure}
\centering
\includegraphics[width=\textwidth]{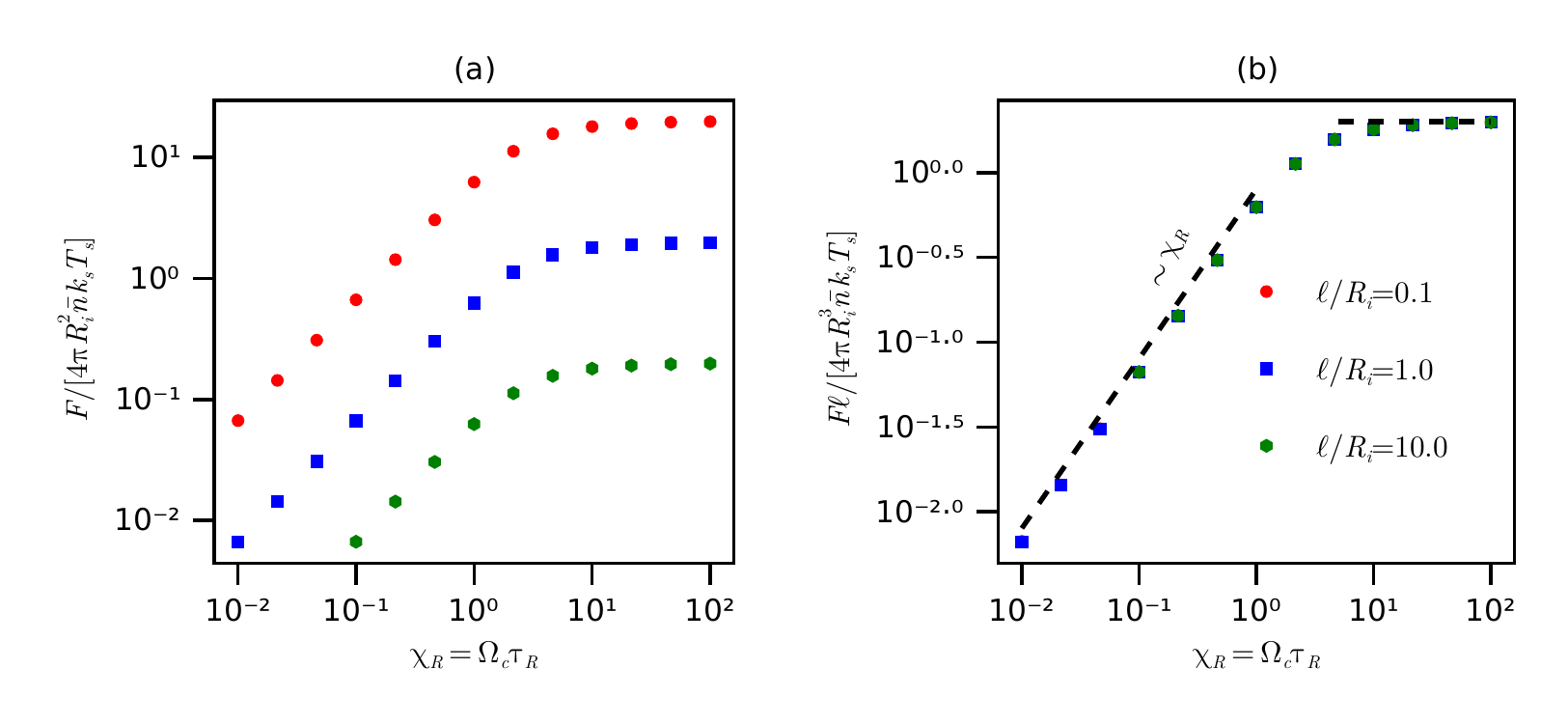}
\caption{\label{fig:field}(a): The magnitude of the dimensionless net force on the interior wall $\bF^w/(4\pi R_i^2\overline{n}\ksts)$  as a function of the field strength $\chi_R$ for different values of $\ell/R_i$. (b): The rescaled net force, $\bF^w \ell /(4\pi R_i^3\overline{n}\ksts)$, as a function of $\chi_R$ for different values of $\ell/R$. All data collapse into one curve in panel (b). The values of $\ell/R_i$ in both panels are the same and are thus only shown in (b). In both panels, the translational diffusion is absent, $D_T\equiv 0$. In the weak-field limit, $\chi_R \ll 1$, the net force is linearly proportional to $\chi_R$ as shown by the dashed line.}
\end{figure}

Because equation \eqref{eq:smol-field} together with its no-flux boundary condition is not analytically tractable, we again make use of Brownian dynamics simulations. In figure \ref{fig:field}(a), we show the dimensionless net force exerted on the interior wall by the ABPs, $\bF^w/(4\pi R_i^2\overline{n}\ksts)$,  as a function of the field strength for different values of $\ell/R$. We note that the net force is in the field direction $\hat{\bH}$. In figure \ref{fig:field}(b), the same data is plotted but with the dimensionless net force multiplied by $\ell/R_i$. This rescaling allows us to collapse all data onto a single curve. In the linear response regime,  the net force is proportional to $\chi_R$. On the other hand, the net force asymptotes to a finite value in the strong field limit. This is due to the fact that at most all $N$ particles are aligned with $\hat{\bH}$ and are pushing against the vesicle; further increasing of the field strength beyond this limit has no effect.

In `wet' active matter systems such as the vesicle problem, the fluid mechanics is ultimately responsible for the motion of the vesicle and needs to be treated properly. Nevertheless, the perspective offered by the dry active matter force balance as discussed in section \ref{sec:intro} gives the right answer for the speed of the vesicle. In particular, consider the case in which the vesicle is driven by an orienting field. The ratio $N^w/N$ is a function of the field strength $\chi_R$, $N^w/N = f(\chi_R)$. As a result, we have the qualitative scaling relation $F^w\sim N \zeta U_s f(\chi_R)$. Noting that $\overline{n} \sim N/R_i^3$ and $\ksts\sim \zeta U_s^2\tau_R$, we have
\begin{equation}
    \frac{F^w}{4\pi R_i^2 \overline{n}\ksts}\sim \frac{ N \zeta U_s}{R_i^2 \overline{n}\ksts} f(\chi_R) \sim \frac{R_i}{\ell}f(\chi_R).
\end{equation}
In the weak-field limit, $f(\chi_R)\sim \chi_R$. For large $\chi_R$, $f(\chi_R) \sim 1$ (independent of $\chi_R$).  The above scaling argument also explains the collapse of the data as shown in figure \ref{fig:field}(b). The maximum that $F^w$ may achieve is $N \zeta U_s$, which gives the result that $F^w\ell/(4\pi R_i^3 \overline{n}\ksts ) = 2$, this is plotted as a horizontal dashed line in figure \ref{fig:field}(b).

We note that in figure \ref{fig:field}, the translational diffusion is absent ($D_T\equiv 0$), which allows the system to achieve the maximum in the net force on the wall. For finite thermal diffusion, the net force is reduced and so does the speed of the vesicle.

\section{Slow variation in activity}
\label{sec:slow-activity-variation}
In the previous section, the dynamics of the vesicle is determined by the distribution of ABPs in the absence of flow. To understand the effect of interior fluid flow on the distribution of ABPs and the dynamics of the vesicle, we consider the case of slow variation in activity. When the activity gradient is small, any smooth variation of the swim speed can be approximated by a Taylor series expansion about the origin. Here, we consider the first effect of a small gradient by keeping the linear term only. The non-dimensional swim speed can be written as
\begin{equation}
    \hat{U}_s(\bx) = 1 + \epsilon \be\bcdot\bx,
\end{equation}
where $\epsilon = |\bnabla U_s| R/U_s \ll 1$ and $\be = \bnabla U_s/|\bnabla U_s|$ is a constant unit vector in the direction of the gradient. If $\epsilon$ is identically zero, we have a spatially homogeneous swim speed and there is no vesicle motion due to spherical symmetry (see discussion in section \ref{sec:conclusion}). In this case of $\epsilon \equiv 0$, the solution is  $\bu_0^\prime = \bu_0^e = \bu_0^s = \bU_0=\bm{0}$, $p_{f,0}^e=0$ and $P_0 = const$. The distribution of ABPs is governed by equations \eqref{eq:smol-no-flow}, \eqref{eq:smol-bc-no-flow} and \eqref{eq:smol-conservation-no-flow} but with $\hat{U}_s = 1$, i.e., this problem reduces to that of ABPs confined inside a fixed spherical domain. This spherical symmetry means that the number density is a function of the radial coordinate only, $n_0(\bx) = n_0(r)$. As shown by \citet{yan_brady_2015}, the number density is a monotonically increasing function that obtains its maximum at the interior wall. Because the total pressure $P_0$ is a constant, this variation of number density (osmotic pressure) maintains a fluid pressure gradient with its maximum at the center of the interior domain. The fluid pressure across the membrane is constant, and no seepage velocity is generated.

To probe the first effect of a small linear gradient, we pose regular expansions for all fields and the translational velocity:
\begin{eqnarray}
    \label{eq:g-expansion-smallgrad}
    g &=& g_0 +\epsilon  g_1 +\cdot\cdot\cdot,\\
    \left(P, p_f^e, p_f^i\right) &=& (P_0, 0, 0) +\epsilon \left(P_1 , p_1^e, p_1^\prime\right)+\cdot\cdot\cdot,\\
    \left(\bu^\prime, \bu^e, \bu^s, \bU\right)&=& \bm{0} + \epsilon \left(\bu_1^\prime ,\bu_1^e, \bu_1^s, \bU_1\right)+\cdot\cdot\cdot.
\end{eqnarray}
At $O(\epsilon)$, the exterior fluid and the interior suspension are still governed by equations \eqref{eq:exterior-momentum}--\eqref{eq:exterior-bc-nondim} and \eqref{eq:interior-momentum}--\eqref{eq:interior-bc-nodim}. Similarly, the seepage velocity is related to the jump in the fluid stress across the membrane given by equation \eqref{eq:non-dimensional-seepage}. The disturbance to the distribution of ABPs at this order is governed by the inhomogeneous equation
\begin{eqnarray}
    \bnabla\bcdot\left( \Pes \bq g_1 - \bnabla g_1\right) - \gamma^2 \nabla_R^2 g_1 &= &- \bnabla\bcdot\left( \alpha Da \bu_1^\prime g_0 + \Pes \be\bcdot\bx \bq g_0\right)\nonumber\\
    &&\label{eq:weak-activity-epsilon}-\frac{1}{2}\alpha Da \bnabla_R\bcdot\left( \bomega_1^\prime g_0\right),
\end{eqnarray}
with the boundary condition
\begin{equation}
    \bn\bcdot\left( \Pes \bq g_1 - \bnabla g_1\right) = -\alpha Da \bn\bcdot\bu_1^\prime g_0 - \Pes \be\bcdot\bx \bn\bcdot\bq g_0\quad\mbox{at}\quad r = \Delta.
\end{equation}
The net disturbance is zero, $\int g_1 d\bx d\bq=0$. As can be seen from equation \eqref{eq:weak-activity-epsilon}, the disturbance fields must be linear to the vector $\be$, which allows us to write the number density in the form
\begin{equation}
    \label{eq:n1-smallgrad}
    n_1 = \be\bcdot\bx h_1(r),
\end{equation}
where $h_1(r)$ is a scalar function of the radial coordinate only.

Due to linearity of the Stokes equations, the interior flow problem at $O(\epsilon)$ admits a solution of the form
\begin{eqnarray}
    P_1 &=& A_1 \be\bcdot\bx,\\
    \bu^\prime_1 &=& A_2 \be + A_3 \be\bcdot\left(\bx\bx - \frac{1}{3}r^2 \bI \right) + \frac{1}{2\beta Da} P_1 \bx.
\end{eqnarray}
Here, the momentum equation \eqref{eq:interior-momentum} is solved using a linear combination of the growing tensor harmonic functions \citep{leal2007advanced}. The continuity equation \eqref{eq:interior-continuity} gives a constraint
\begin{equation}
    \label{eq:interior-continuity-coef}
    5 A_3 + \frac{3 A_1}{\beta Da}=0.
\end{equation}

We can solve the external flow problem by considering two separate problems with different boundary conditions: (1) $\bu^e_1 = \bu_1^s$ and (2) $\bu^e_1 = \bU_1$ at $r=1$. Instead of solving the flow field due to the second boundary condition in terms of the yet unknown velocity $\bU_1$, it will be determined from the reciprocal theorem \eqref{eq:reciprocal-dimensionless}. As a result, one only needs to compute the exterior flow field due to the seepage velocity $\bu_1^s$. The exterior flow problem with the first boundary condition  has a solution of the form
\begin{eqnarray}
    p_1^e &=& A_4 \be\bcdot\frac{\bx}{r^3},\\
    \label{eq:u1e-form}
    \bu^e_1 &=& A_5 \be \frac{1}{r} + A_6 \be\bcdot\left( \frac{\bI}{r^3} -3 \frac{\bx\bx}{r^5} \right) + \frac{1}{2 Da} p_1^e \bx,
\end{eqnarray}
where the decaying tensor harmonic functions are used.
To satisfy the continuity equation \eqref{eq:exterior-continuity}, we must have
\begin{equation}
    \label{eq:exterior-continuity-coef}
    A_4 = 2 Da A_5.
\end{equation}
The seepage velocity connects the interior and exterior flow field via
\begin{equation}
    \bu_1^\prime(\bx = \Delta \be_r) = \bu_1^s = \bu_1^e(\bx = \be_r),
\end{equation}
which reduces to
\begin{equation}
    \label{eq:seepage-connect-coef}
    A_2 - \frac{1}{3}\Delta^2 A_3=A_5+A_6\quad\mbox{and}\quad A_3\Delta^2 + \frac{A_1\Delta^2}{2\beta Da} = -3 A_6 + \frac{A_4}{2Da}.
\end{equation}

The volume conservation \eqref{eq:volume-conservation} is satisfied. The velocity of the vesicle is obtained from the reciprocal theorem, which gives
\begin{equation}\label{eq:U1-reciprocal}
    \bU_1 = -\frac{1}{4\pi}\int_{S^2} \bu_1^s d\Omega  =-\left(A_2 + \frac{A_1\Delta^2}{6\beta Da}\right) \be.
\end{equation}

Finally, to solve equation \eqref{eq:non-dimensional-seepage} at this order, we need to compute the fluid stress at the interior and the exterior wall. At the interior wall, we have
\begin{eqnarray}
    \bsigma_{f,1}^i\bcdot\be_r  =&& -\Delta \left( A_1 - h_1(\Delta) \frac{\kt}{\ksts}\right)\be\bcdot\be_r\be_r \nonumber\\
    &&+\Delta \left(\frac{7}{3}A_3 \beta Da + \frac{3}{2}A_1\right)\be\bcdot\be_r\be_r  + \Delta\left( \frac{1}{3}A_3 \beta Da + \frac{1}{2}A_1\right)\be.
\end{eqnarray}
The traction at the exterior wall has two contributions. The first is due to the vesicle translating at a constant speed $\bU_1$, which is given by \citep[pp. 44]{guazzelli2011physical}
\begin{equation}
    \bsigma_{U_1}^e\bcdot\be_r  = -\frac{3}{2}Da \bU_1.
\end{equation}
The second contribution is from the seepage velocity boundary condition $\bu_1^s$, which is given by
\begin{equation}
    \label{eq:traction-u1s}
    -A_4 \be\bcdot\be_r\be_r + Da\left(-A_5 -6A_6 +\frac{A_4}{2Da} \right)\be + Da\left(-A_5 +18A_6 -\frac{3A_4}{2Da} \right)\be\bcdot\be_r\be_r.
\end{equation}
Using equations \eqref{eq:U1-reciprocal}--\eqref{eq:traction-u1s} we can obtain the jump in the fluid stress across the membrane, which then allows us to calculate the seepage velocity using equation \eqref{eq:non-dimensional-seepage}. Equating this result with the seepage velocity obtained from equation \eqref{eq:u1e-form} by setting $r=1$, we arrive at the following equations for the coefficients:
\begin{equation}\label{eq:A5-A6}
    A_5+A_6 = 0,
\end{equation}
and
\begin{eqnarray}
    \label{eq:seepage-coef-2}
    \frac{A_4}{2Da}-3A_6=&&A_1\Delta \left(-1 + \frac{\Delta}{4\beta} \right) + \frac{3}{2}Da A_2-\frac{8}{3}  A_3 \beta  \Delta Da-2A_4\nonumber\\
    && -2A_5Da+12A_6Da - \Delta h_1(\Delta)\frac{\kt}{\ksts}.
\end{eqnarray}
Equation \eqref{eq:A5-A6} implies that $\bu_1^s$ is proportional to $\be\bcdot\be_r\be_r$ and the component proportional to $\be$ is zero, which is consistent with the fact that the seepage velocity is in the normal ($\be_r$) direction. At this stage, we have obtained 6 equations for the 6 unknown coefficients $A_i$ ($i=1\cdot\cdot\cdot 6$), which are given by equations \eqref{eq:interior-continuity-coef}, \eqref{eq:exterior-continuity-coef}, \eqref{eq:seepage-connect-coef}, \eqref{eq:A5-A6} and \eqref{eq:seepage-coef-2}. Using these equations, one could express $A_i$ in terms of the boundary value of $h_1$ at the interior wall, i.e., $h_1(\Delta)$. These relations are obtained as
\begin{equation}
    A_6 = \frac{\Delta^2}{4} \frac{k_BT}{k_sT_s} \frac{h_1(\Delta)}{\Delta + Da (6 \beta + 4 \Delta)},
\end{equation}
and
\begin{eqnarray}
    A_1 &=& \frac{40 Da \beta}{\Delta^2}A_6,\quad A_2 = -8 A_6, \quad A_3 = -\frac{24}{\Delta^2}A_6,\\
    A_4 &=& -2Da A_6,\quad A_5 = -A_6.
\end{eqnarray}

From equation \eqref{eq:U1-reciprocal}, we have the net motion of the vesicle
\begin{equation}
    \label{eq:U1-smallgrad-solution}
    \bU_1 = \frac{4}{3}A_6\be=\frac{\Delta^2}{3} \frac{k_BT}{k_sT_s} \frac{h_1(\Delta)}{\Delta + Da (6 \beta + 4 \Delta)}\be.
\end{equation}
Equation \eqref{eq:U1-smallgrad-solution} is the main result of this section. In obtaining \eqref{eq:U1-smallgrad-solution} the only assumption made is the small linear gradient in the swim speed; therefore, it applies generally for all ranges of the parameters $\alpha$, $\beta$, $Da$, $\Pes$ and $\gamma$. In particular, no restriction on the activity of the ABPs (e.g., $\ell/\delta$) is made. We note that $h_1(\Delta)$ depends parametrically on all the above parameters.

To obtain $h_1(r)$, we need to solve equation \eqref{eq:weak-activity-epsilon} that governs the disturbance probability density distribution of the ABPs. As an approximation, we consider the general solution using the $\bQ=\bm{0}$ closure. At $O(1)$, the spherical symmetry allows us to write the number density and polar order in the form
\begin{eqnarray}
    n_0(\bx) &=& n_0(r),\\
    \boldm_0(\bx) &=& \bx f(r),
\end{eqnarray}
which, when inserted into equations \eqref{eq:number-equation} and \eqref{eq:polar-equation}, leads to a couple of ordinary differential equations (ODEs) for $n_0(r)$ and $f(r)$. The solutions to $n_0$ and $\boldm_0$ under this assumption are obtained by \cite{yan_brady_2015}.

Next, we consider the disturbance distribution of ABPs at $O(\epsilon)$. At this order, the number density distribution is governed by
\begin{equation}
\label{eq:number-smallgrad}
     \bnabla\bcdot\bj_{n,1}=0 \quad\mbox{and}\quad \bj_{n,1} =\alpha Da \bu_1^\prime n_0 + \Pes \boldm_1 +\Pes \be\bcdot\bx \boldm_0 - \bnabla n_1.
\end{equation}
The no-flux boundary condition is $\bn\bcdot\bj_{n,1} =0$ at $r = \Delta$. Similarly, the governing equation for the polar order (assuming $\bQ_1=\bm{0}$) is
\begin{equation}
\label{eq:polar-smallgrad}
    \bnabla\bcdot\bj_{m,1} +2 \gamma^2 \boldm_1 - \frac{1}{2}
    \alpha Da \bomega_1^\prime\times\boldm_0=0,
\end{equation}
and
\begin{equation}
\label{eq:polar-flux-smallgrad}
    \bj_{m,1} = \alpha Da \bu_1^\prime \boldm_0 +\frac{1}{3}\Pes \left(n_0 \be\bcdot\bx + n_1 \right) \bI- \bnabla\boldm_1.
\end{equation}
 No-flux at $r = \Delta$ is  $\bn\bcdot\bj_{m,1}=\bm{0}$. Similar to equation \eqref{eq:n1-smallgrad}, linearity and symmetry allows us to write the solution to the polar order in the form
\begin{equation}
    \label{eq:m1-smallgrad}
    \boldm_1 = \be h_2(r) + \be\bcdot\bx\bx h_3(r),
\end{equation}
where $h_2(r)$ and $h_3(r)$ are functions of the radial coordinate only and satisfy a coupled set of ODEs that can be found in Appendix \ref{sec:appendix-smallgrad}.

\begin{figure}
    \centering
    \includegraphics[width=0.7\textwidth]{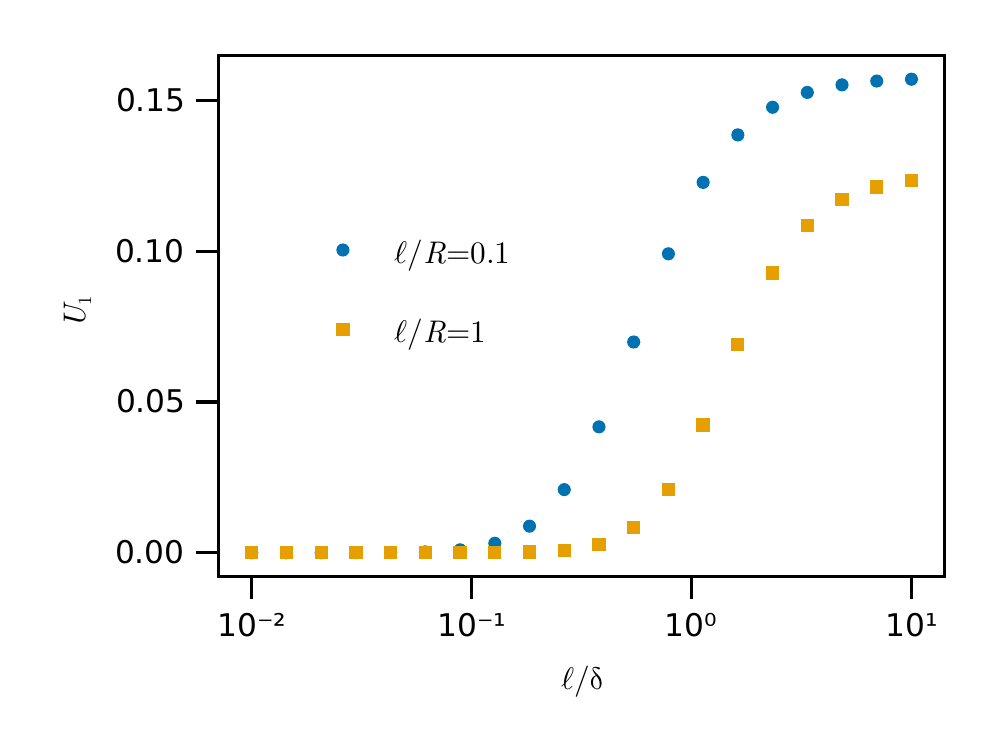}
    \caption{\label{fig:slow-variation}The dimensionless speed of the vesicle $U_1$ as a function of $\ell/\delta$ for different fixed values of $\ell/R$. All other parameters are fixed: $\Delta=0.98, \alpha=1, Da=0.1$ and $\beta=1.0$. }
\end{figure}

In figure \ref{fig:slow-variation} we show the dimensionless speed of the vesicle ($U_1$) as a function of $\ell/\delta$ for $\ell/R = \{0.1, 1\}$. With other dimensionless parameters fixed, the increase of $\ell/\delta$ means the decrease of the translational diffusivity and thus the increase of activity. The speed of the vesicle vanishes as the activity approaches zero, $\ell/\delta \to 0$. As $\ell/\delta$ increases, the speed of the vesicle increases and asymptotes to a finite value for large $\ell/\delta$. The speed is larger for a smaller $\ell/R$ because a thin boundary layer near the interior wall develops that enhances the front-back asymmetry of the density distribution.

\section{Concluding remarks}
\label{sec:conclusion}
In this paper we have proposed a composite low-Reynolds-number propulsion system made up of  active Brownian particles encapsulated in a vesicle for the purpose of enhanced transport beyond that of passive Brownian diffusion. Instead of using the self-propulsion of a microswimmer directly, such as by attaching a cargo to its surface, we considered an alternate mechanism in which the vesicle is propelled by a fluid seepage velocity generated by a concentration gradient of these encapsulated particles. In the present work, we considered the cases in which the concentration gradient is generated by either a prescribed activity gradient in the swim speed of these ABPs or an external orienting field. By tuning the spatial pattern of variation in the swim speed, one could obtain a concentration profile that in turn propels the vesicle with a certain speed or in a desired direction. Alternatively, the application of an external orienting field can push the ABPs against the wall and generate net thrust for the vesicle. We provided a continuum formulation governing the dynamics of the vesicle-ABPs system and explicitly analyzed its behavior in the limits of weak interior flow and small activity gradient. For the composite system as a whole, it moves by jet propulsion at low Reynolds number, i.e.,  fluid is drawn in from one side of the vesicle and expels from the other. The encapsulation of ABPs only provides a mechanism to generate such a seepage flow.

We emphasize that in the present model it is the concentration gradient rather than the species of the solute particles that is ultimately responsible for vesicle locomotion. Any osmotic solute, not necessarily active, is able to propel the vesicle so long as a concentration gradient is maintained. For a passive solute, one can maintain a concentration gradient using chemical reaction, e.g., by placing a distribution of sources and sinks. In this paper, we analyzed how such a concentration gradient may be generated by an activity gradient or by the application of an external orienting field. For magnetotactic bacteria or synthetic active particles, an aligning magnetic field can be used to control the direction of the concentration gradient and therefore the direction of motion of the vesicle.

In an experimental setting, a spatial variation of the swim speed  of photokinetic bacteria can be achieved by exposing the bacteria to external light intensity gradients. These light-powered bacteria exhibit a larger swim speed in regions of higher light intensity.  Under spatially patterned light fields, light-responsive bacteria can self-assemble into reconfigurable structures---`painting' with bacteria \citep{arlt2018painting,elife_painting}. Another possible mechanism for inducing a spatially varying swim speed could be the spatial modulation of `fuel' (food sources).

For magnetotactic bacteria, instead of spatial modulation of swim speed one can use an external static magnetic field that tends to align the bacteria in a certain direction.  For static or slowly-varying magnetic fields, the magnitude of the induced electric field in this low frequency limit ($\ll 100$ kHz) is small so that its effect on the membrane dynamics is negligible \citep{ye2015vesicle}.

In obtaining the results we assumed that the ABPs can be treated as a continuum and only contribute to the suspension stress via the osmotic pressure. We note that additional constitutive models at the continuum level for the suspension stress can be readily incorporated into our model. The hydrodynamic interactions of the active particles with each other or the confining vesicle boundary is neglected. These effects can be studied using a colloidal approach by considering the detailed interactions among the active particles and with the boundary. For example, this is considered in the study of a single squirmer encapsulated in a porous container by \citet{marshall_brady_2021} and for the case of a collection of squirmers inside a droplet that is immersed in another fluid by \citet{Huang2020}.

To achieve net motion of the spherical vesicle, a number density distribution at the vesicle interior wall that breaks the front-back symmetry is required. Instead of maintaining an asymmetric density distribution in a spherical vesicle using ABPs with spatially-varying properties or ABPs with constant properties but in an orienting field, one can also consider an asymmetric vesicle. For ABPs with constant properties confined in an asymmetric container, a symmetry-breaking density distribution will emerge because the accumulation of ABPs at the wall depends on the local curvature. The effect of vesicle shape on its net motion is left for a future study.

The enhancement of transport revealed by our study may be useful for the development of synthetic microscale propelling systems for the purpose of delivery of therapeutic payloads, penetrating complex media, or clearing clogged arteries. We hope that our proposed theoretical designs can prompt new experimental implementations.

\section*{Funding}
This work is supported by the National Science Foundation under Grant No. CBET 1803662.

\appendix

\section{Brownian dynamics simulations}
\label{sec:BD}
The dynamics of ABPs confined inside the vesicle in an external orienting field can be resolved using Brownian dynamics (BD) simulations. Each ABP follows the Langevin equations of motion given by
\begin{equation}
    \bm{0} = -\zeta\left(\bU-U_s \bq \right) +\bF^B +\bF^w \quad\mathrm{and}\quad \bm{0}=-\zeta_R \bOmega + \bL^B +\bL^e,
\end{equation}
where $\bU$ ($\bOmega$) is the instantaneous linear (angular) velocity, $\bF^B$ is the Brownian force, $\bF^w$ is the hard-sphere force due to collisions with the interior wall, $\zeta_R$ is the rotary Stokes drag coefficient, $\bL^B$ is the Brownian torque and $\bL^e$ is the external torque due to the field.

The Brownian force and torque satisfy the white noise statistics: $\overline{\bF^B}=\bm{0}, \overline{\bF^B(0)\bF^B(t)}=2\kt \zeta \delta(t)\bI$, and $\overline{\bL^B}=\bm{0}, \overline{\bL^B(0)\bL^B(t)}=2 \zeta_R^2 \delta(t)\bI/\tau_R$. Here, $\delta(t)$ is the delta function. In the BD simulations, the particle orientations are represented using unit quaternions. At each time step, the instantaneous particle velocities are computed and then used to update the positions and orientations. The kinematic equation relating the angular velocity and the rate-of-change of the quaternion is given by \citet{delong2015}.

In figure \ref{fig:field}, all data points are obtained by averaging over the long-time behavior of the system. In each simulation, $10^5$ noninteracting ABPs are used, and the system is evolved for a sufficiently long time such that the steady state is reached.

\section{Equations for $h_1, h_2$ and $h_3$}
\label{sec:appendix-smallgrad}

In this appendix we provide the detail on the derivation of the ODEs for $h_1, h_2$ and $h_3$. Note that the conservation
\begin{equation}
    \int_{ |\bx| \leq \Delta} n_1d\bx =0
\end{equation}
is satisfied.

Note that
\begin{equation}
    \bnabla f(r) =  \bx \frac{1}{r}f^\prime,
\end{equation}
and
\begin{equation}
    \bnabla(\be\bcdot\bx f) = \be f + \be\bcdot\bx\bx \frac{1}{r}f^\prime.
\end{equation}

Using  the identity
\begin{equation}
    \bnabla\bcdot(\underbrace{\bx\bx\cdot\cdot\cdot\bx}_{k}f(r)) = \left[(d+k-1)f+ r f^\prime\right]\underbrace{\bx\bx\cdot\cdot\cdot\bx}_{k-1},
\end{equation}
we can obtain
\begin{eqnarray}
    \bnabla\bcdot[\be\bcdot\bx\bx f(r)] &=& \be\bcdot[\bnabla\bcdot(\bx\bx f(r))]= \be\bcdot\bx (4f + r f^\prime),\\
    \bnabla\bcdot[\be\bcdot\bx\bx\bx f(r)] &=& \be\bcdot[\bnabla\bcdot(\bx\bx\bx f(r))]= \be\bcdot\bx\bx (5f + r f^\prime).\\
\end{eqnarray}
Similarly, we have
\begin{equation}
    \nabla^2 f =  \frac{2}{r}f^\prime + f^{\prime\prime},
\end{equation}
\begin{equation}
    \nabla^2[\be\bcdot\bx f] = \be\bcdot\bx \left( \frac{4 f^\prime}{r} + f^{\prime\prime}\right),
\end{equation}
\begin{equation}
    \nabla^2(\be\bcdot\bx\bx f(r)) = 2 \be f + \be\bcdot\bx\bx \left(\frac{6}{r}f^\prime +f^{\prime\prime} \right).
\end{equation}
The equation for $h_1$ is given by
\begin{eqnarray}
    && \alpha Da \frac{d n_0 }{d r} \left( \frac{1}{r}A_2 + \frac{2}{3}r A_3 + \frac{r}{2\beta Da}A_1\right) + \Pes \left(\frac{1}{r}\frac{d h_2}{dr} + 4h_3 +r \frac{dh_3}{dr} \right)\nonumber \\
    && +\Pes \left(4 f + r \frac{df}{dr}\right) - \frac{4}{r}\frac{d h_1}{dr} - \frac{d^2 h_1}{dr^2}=0.
\end{eqnarray}

The no-flux condition is given by
\begin{equation}
    \Pes (r^2 f + h_2 + r^2 h_3) - h_1 - r \frac{dh_1}{dr} +A_2 \alpha Da n_0 + \frac{\alpha}{6\beta} r^2 n_0 (3A_1+4A_3Da \beta)=0,
\end{equation}
evaluated at $r=\Delta$. The governing equation for $h_2$ is
\begin{eqnarray}
    &&\alpha Da \left(A_2 - \frac{1}{3}r^2 A_3 \right)f + \frac{1}{3}\Pes \left( n_0 + h_1\right) - \frac{2}{r}\frac{dh_2}{dr} - \frac{d^2h_2}{dr^2}-2h_3 \nonumber\\
    &&+ 2\gamma^2 h_2 +\frac{1}{2}\alpha Da \left(\frac{5}{3}A_3 +\frac{A_2}{2\beta Da}\right)r^2f=0.
\end{eqnarray}

The no-flux condition at $r = \Delta$ is
\begin{equation}
    \frac{d h_2}{d r}=0.
\end{equation}

The governing equation for $h_3$ is
\begin{eqnarray}
    &&\alpha Da \left(A_3 +\frac{A_1}{2\beta Da} \right)f + \alpha Da \frac{1}{r}\frac{df}{dr}\left( A_2 + \frac{2}{3}r^2A_3 + \frac{A_1r^2}{2\beta Da}\right)\nonumber\\
    &&+\frac{1}{3}\Pes \frac{1}{r}\left(\frac{dn_0}{dr} + \frac{dh_1}{dr} \right) - \frac{6}{r}\frac{dh_3}{dr} - \frac{d^2h_3}{dr^2}\nonumber \\
    &&+2\gamma^2 h_3-\frac{1}{2}\alpha Da \left(\frac{5}{3}A_3 +\frac{A_2}{2\beta Da}\right)f=0.
\end{eqnarray}
The no-flux condition is
\begin{equation}
    \alpha Da r f \left(A_2 +\frac{2}{3}r^2 A_3 + \frac{r^2 A_1}{2\beta Da} \right) +\frac{1}{3}\Pes r (n_0+h_1) - 2r h_3 - r^2 \frac{dh_3}{dr}=0,
\end{equation}
evaluated at $r=\Delta$. We solve these equations in MATLAB using a Chebyshev collocation method \citep{Trefethen}.

\bibliography{reference}

\end{document}